\documentclass[reprint,onecolumn,notitlepage,showpacs,preprintnumbers,nofootinbib,amsmath,amssymb,aps,prd,floatfix,superscriptaddress]{revtex4-1}
\pdfoutput=1
\usepackage{graphicx}
\usepackage{bm}
\usepackage{url}
\usepackage[hyperindex]{hyperref}
\usepackage{color}
\usepackage[ddmmyy,24hr]{datetime}
\usepackage{bigdelim}
\usepackage{booktabs}
\usepackage{dcolumn}
\usepackage{multirow}
\usepackage[large]{subfigure}
\usepackage{cancel}
\usepackage{stackrel}
\usepackage{paralist}
\usepackage{xspace}
\usepackage{slashed}
\newcommand{\nua}[1]{\ensuremath{\rlap{\kern-2.5pt\ensuremath{\overset{\scriptscriptstyle(-)}{\phantom{\nu}}}}{\ensuremath{{\nu}_{#1}}}}}
\newcommand{\vet}[1]{\ensuremath{\hskip-1pt\vec{\hskip1pt#1}}}
\usepackage{enumitem}

\newcommand{\cenns}{CE$\nu$NS\xspace}

\begin{document}

\title{Physics results from the first COHERENT observation of CE$\nu$NS in argon and their combination with cesium-iodide data}

\author{M. Cadeddu}
\email{matteo.cadeddu@ca.infn.it}
\affiliation{Istituto Nazionale di Fisica Nucleare (INFN), Sezione di Cagliari,
Complesso Universitario di Monserrato - S.P. per Sestu Km 0.700,
09042 Monserrato (Cagliari), Italy}

\author{F. Dordei}
\email{francesca.dordei@cern.ch}
\affiliation{Istituto Nazionale di Fisica Nucleare (INFN), Sezione di Cagliari,
Complesso Universitario di Monserrato - S.P. per Sestu Km 0.700,
09042 Monserrato (Cagliari), Italy}

\author{C. Giunti}
\email{carlo.giunti@to.infn.it}
\affiliation{Istituto Nazionale di Fisica Nucleare (INFN), Sezione di Torino, Via P. Giuria 1, I--10125 Torino, Italy}

\author{Y.F. Li}
\email{liyufeng@ihep.ac.cn}
\affiliation{Institute of High Energy Physics,
Chinese Academy of Sciences, Beijing 100049, China}
\affiliation{School of Physical Sciences, University of Chinese Academy of Sciences, Beijing 100049, China}

\author{E. Picciau}
\email{emmanuele.picciau@ca.infn.it}
\affiliation{Dipartimento di Fisica, Universit\`{a} degli Studi di Cagliari,
and
INFN, Sezione di Cagliari,
Complesso Universitario di Monserrato - S.P. per Sestu Km 0.700,
09042 Monserrato (Cagliari), Italy}
\affiliation{University of Massachusetts, Amherst (MA), 01003, USA}

\author{Y.Y. Zhang}
\email{zhangyiyu@ihep.ac.cn}
\affiliation{Institute of High Energy Physics,
Chinese Academy of Sciences, Beijing 100049, China}
\affiliation{School of Physical Sciences, University of Chinese Academy of Sciences, Beijing 100049, China}

\date{3 July 2020}

\begin{abstract}
We present the results
on the radius of the neutron distribution in $^{40}\text{Ar}$,
on the low-energy value of the weak mixing angle,
and on the electromagnetic properties of neutrinos
obtained from the analysis of the
coherent neutrino-nucleus elastic scattering data in argon recently published by
the COHERENT collaboration, taking into account proper radiative corrections.
We present also the results of the combined analysis of the
COHERENT argon and cesium-iodide data
for the determination of
the low-energy value of the weak mixing angle
and the electromagnetic properties of neutrinos.
In particular,
the COHERENT argon data allow us to improve significantly the only existing
laboratory bounds on the electric charge $q_{\mu\mu}$ of the muon neutrino
and on the transition electric charge $q_{\mu\tau}$.
\end{abstract}


\maketitle

\section{Introduction}
\label{sec:introduction}

The observation of coherent elastic neutrino-nucleus scattering (\cenns) in cesium-iodide performed in 2017 by the COHERENT experiment~\cite{Akimov:2017ade,Akimov:2018vzs} unlocked an innovative and powerful tool to study many and diverse physical phenomena~\cite{Cadeddu:2017etk,Papoulias:2019lfi,Coloma:2017ncl,Liao:2017uzy,Kosmas:2017tsq,Denton:2018xmq,AristizabalSierra:2018eqm,Cadeddu:2018dux,Dutta:2019eml,Dutta:2019nbn}.
Recently, the COHERENT collaboration observed \cenns for the first time also in argon~\cite{Akimov:2020pdx}, using a single-phase 24 kg liquid-argon (LAr) scintillation detector, with two independent analyses that prefer \cenns over the background-only null hypothesis with greater  than  $3\,\sigma$ significance. The experimental challenge behind this analysis is the need to observe
nuclear recoils with a very small kinetic energy $T_\mathrm{nr}$ of a few keV, and thus the need of a low nuclear-recoil-energy threshold, in presence of a larger background, when compared to the cesium-iodide case.
This requirement is necessary for the coherent recoil of the nucleus
which occurs for 
$|\vec{q}| R \ll 1$~\cite{Bednyakov:2018mjd},
where $|\vec{q}| \simeq \sqrt{2 M T_\mathrm{nr}}$ is the three-momentum transfer,
$R$ is the nuclear radius of a few fm,
and
$M$ is the nuclear mass,
of about 40 GeV for argon nuclei. The observation in argon, which is the lightest nucleus for which \cenns process has been measured, allows to demonstrate the \cenns cross-section dependence on the square of the number of neutrons $N^2$, but it can also provide valuable information on nuclear physics, neutrino properties, physics beyond the standard model (SM), and electroweak (EW) interactions. 

In this paper, we present the bounds on different parameters of the EW interaction and neutrino electromagnetic properties obtained analyzing the
new COHERENT Ar data and those obtained with a combined analysis of the
COHERENT CsI and Ar data,
using the results of the analysis of the CsI data in Ref.~\cite{Cadeddu:2019eta}.
During the completion of this work,
another analysis of this type appeared on arXiv~\cite{Miranda:2020tif},
but the results are not comparable with ours because
we fit the COHERENT Ar data,
whereas the analysis of Ref~\cite{Miranda:2020tif}
is not a fit of the COHERENT Ar data,
but a fit of the number of CE$\nu$NS events obtained by the
COHERENT collaboration from the fit of the data~\cite{Akimov:2020pdx}.
Such an indirect analysis
underestimates the systematic uncertainties,
especially those due to the background that are not taken into account.

The plan of the paper is as follows.
In Section~\ref{sec:theory} we describe briefly the \cenns formalism used in our calculation as well as the inputs employed for simulating the signal spectra.
In Sections~\ref{sec:neutron} and \ref{sec:electroweak}
we derive, respectively, the results on
the average rms radius of the neutron distributions in Ar
and
on the weak mixing angle.
In Sections~\ref{sec:radii}, \ref{sec:charges}, and \ref{sec:magnetic}
we present, respectively,
the constraints on the neutrino charge radii,
neutrino electric charges and magnetic moments.
Finally, in Section~\ref{sec:conclusions} we summarize the results of the paper.

\section{Formalism and signal predictions}
\label{sec:theory}

The SM weak-interaction differential cross section as a function of the nuclear kinetic recoil energy $T_\mathrm{nr}$
of \cenns processes with a spin-zero nucleus $\mathcal{N}$ with $Z$ protons and $N$ neutrons is given by~\cite{Drukier:1983gj,Barranco:2005yy,Patton:2012jr}
\begin{equation}
\dfrac{d\sigma_{\nu_{\ell}\text{-}\mathcal{N}}}{d T_\mathrm{nr}}
(E,T_\mathrm{nr})
=
\dfrac{G_{\text{F}}^2 M}{\pi}
\left(
1 - \dfrac{M T_\mathrm{nr}}{2 E^2}
\right)
\left[
g_{V}^{p}
Z
F_{Z}(|\vet{q}|^2)
+
g_{V}^{n}
N
F_{N}(|\vet{q}|^2)
\right]^2
,
\label{cs-std}
\end{equation}
where $G_{\text{F}}$ is the Fermi constant, $\ell = e, \mu, \tau$ is the neutrino flavor, $E$ is the neutrino energy.
The well-known tree-level values of
$g_{V}^{p}$
and
$g_{V}^{n}$
are
\begin{equation}
g_{V}^{p}
=
\dfrac{1}{2} - 2 \sin^2\!\vartheta_{W}
,
\qquad
g_{V}^{n}
=
- \dfrac{1}{2}
,
\label{gV}
\end{equation}
where
$\vartheta_{\text{W}}$ is the weak mixing angle, also known as the Weinberg angle.
In this paper we consider the following more accurate values that take into account
radiative corrections in the $\overline{\text{MS}}$ scheme~\cite{Erler:2013xha}:
\begin{align}
\null & \null
g_{V}^{p}(\nu_{\ell})
=
\rho \left( \dfrac{1}{2} - 2 \sin^2\!\vartheta_{W} \right)
- \dfrac{ \hat{\alpha}_{Z} }{ 4 \pi \hat{s}^2_{Z} }
\left( 1 - 2 \, \dfrac{ \hat{\alpha}_{s}(m_{W}) }{ \pi } \right)
+ \dfrac{ \alpha }{ 6 \pi } \left( 3 - 2 \ln\dfrac{m_{\ell}^2}{m_W^2} \right)
,
\label{gVpth}
\\
\null & \null
g_{V}^{n}
=
- \dfrac{\rho}{2}
- \dfrac{ \hat{\alpha}_{Z} }{ 8 \pi \hat{s}^2_{Z} }
\left( 7 - 5 \, \dfrac{ \hat{\alpha}_{s}(m_{W}) }{ \pi } \right)
,
\label{gVnth}
\end{align}
where
\begin{equation}
\sin^2\!\vartheta_{W}
=
0.23857 \pm 0.00005
\,
\text{\protect\cite{Tanabashi:2018oca}}
\label{s2tw_0}
\end{equation}
is the low-energy value of the weak mixing angle,
often denoted with $\hat{s}^2_{0}$~\cite{Erler:2013xha,Tanabashi:2018oca},
and
\begin{align}
\null & \null
\rho
=
1.00058
\,
\text{\protect\cite{Tanabashi:2018oca}}
,
\\
\null & \null
\hat{s}^2_{Z}
=
0.23122 \pm 0.00003
\,
\text{\protect\cite{Tanabashi:2018oca}}
,
\label{s2tw_Z}
\\
\null & \null
\hat{\alpha}_{Z}^{-1}
=
127.955 \pm 0.010
\,
\text{\protect\cite{Tanabashi:2018oca}}
,
\label{alpha_em_Z}
\\
\null & \null
\hat{\alpha}_{s}(m_{W})
=
0.123 \pm 0.018 \pm 0.017
\,
\text{\protect\cite{Alitti:1991yh}}
\label{alpha_strong_W}
\end{align}
are, respectively,
the $\rho$ parameter of electroweak interactions,
the value of $\sin^2\!\vartheta_{W}$
at the scale of the $Z$-boson mass,
the value of the electromagnetic fine-structure constant
at the scale of the $Z$-boson mass,
and
the value of the strong constant
at the scale of the $W$-boson mass.
The value of $\hat{\alpha}_{s}(m_{W})$ in Eq.~(\ref{alpha_strong_W})
is the only measured one that we found in the literature.
It is in agreement with the PDG summary in Figure~9.5 of Ref.~\cite{Tanabashi:2018oca}.
In any case, a precise value of $\hat{\alpha}_{s}(m_{W})$
is not needed, because its contribution is practically negligible.

The terms in Eqs.~(\ref{gVpth}) and (\ref{gVnth})
proportional to
$\hat{\alpha}_{Z} / \hat{s}^2_{Z}$,
which in turn is proportional to the square of the charged-current weak coupling constant,
are due to box diagrams with $W$-boson propagators.
The last term in Eq.~(\ref{gVpth}) depends on the flavor $\ell$ of the interacting
neutrino $\nu_{\ell}$ through the
corresponding charged lepton mass $m_{\ell}$.
This term can be interpreted as the contribution of the neutrino charge radius
and is consistent with the expression of the neutrino charge radius
calculated in Refs.~\cite{Bernabeu:2000hf,Bernabeu:2002nw,Bernabeu:2002pd},
that we will discuss in Section~\ref{sec:radii}.

Numerically,
neglecting the small uncertainties,
we obtain
\begin{align}
\null & \null
g_{V}^{p}(\nu_{e})
=
0.0401
,
\label{gVp-nue}
\\
\null & \null
g_{V}^{p}(\nu_{\mu})
=
0.0318
,
\label{gVp-num}
\\
\null & \null
g_{V}^{n}
=
-0.5094
.
\label{gVn}
\end{align}
These values are different from the tree-level values
$g_{V}^{p}=0.0229$
and
$g_{V}^{n}=-0.5$
obtained with Eq.~(\ref{gV}),
especially those of $g_{V}^{p}(\nu_{e})$ and $g_{V}^{p}(\nu_{\mu})$.

In Eq.~(\ref{cs-std})
$F_{Z}(|\vet{q}|^2)$
and
$F_{N}(|\vet{q}|^2)$
are, respectively, the form factors of the proton and neutron distributions in the nucleus.
They are given by the Fourier transform of the corresponding nucleon
distribution in the nucleus and
describe the loss of coherence for
$|\vet{q}| R_{p} \gtrsim 1$
and
$|\vet{q}| R_{n} \gtrsim 1$,
where $R_{p}$ and $R_{n}$ are, respectively, the rms radii of the proton and neutron distributions.
For the two form factors one can use different parameterizations.
The three most popular ones are the
symmetrized Fermi~\cite{Piekarewicz:2016vbn},
Helm~\cite{Helm:1956zz}, and
Klein-Nystrand~\cite{Klein:1999qj}
parameterizations that give practically identical results,
as we have verified (see Figure~\ref{fig:rn-chi}).
Here, we briefly describe only the Helm parameterization
(descriptions of the other parameterizations can be found in
several papers,
for example in Refs.~\cite{Piekarewicz:2016vbn,Cadeddu:2017etk,Khan:2019mju,Papoulias:2019txv}),
that is given by
\begin{equation}
F^{\text{Helm}}(q^2)
=
3
\,
\dfrac{j_{1}(q R_{0})}{q R_{0}}
\,
e^{- q^2 s^2 / 2}
,
\label{ffnHelm}
\end{equation}
where
$
j_{1}(x) = \sin(x) / x^2 - \cos(x) / x
$
is the spherical Bessel function of order one
and $R_{0}$ is the box (or diffraction) radius.
The rms radius $R$ of the corresponding nucleon distribution is given by
\begin{equation}
R^2 = \dfrac{3}{5} \, R_{0}^2 + 3 s^2
.
\label{RnHelm}
\end{equation}
For the parameter $s$, that quantifies the so-called surface thickness, we consider the value $s = 0.9 \, \text{fm}$
which was determined for the proton form factor of similar nuclei
\cite{Friedrich:1982esq}.

We determined the value of
the rms proton distribution radius $R_{p}$ from the value of the
$^{40}\text{Ar}$ charge radius measured precisely
in electromagnetic experiments~\cite{Fricke:1995zz,Angeli:2013epw}:
\begin{equation}
R_{c}
=
3.4274 \pm 0.0026 \, \text{fm}.
\label{Rc-exp}
\end{equation}
The charge radius $R_{c}$ is given
by~\cite{Ong:2010gf,Horowitz:2012tj}\footnote{
Other contributions considered in Refs.~\cite{Ong:2010gf,Horowitz:2012tj}
are negligible.
They are
the Darwin-Foldy contribution
$ 3 / 4 M^2 \simeq 0.033 \, \text{fm}^2 $,
and the spin-orbit charge density contribution
$ \langle r^2 \rangle_{\text{so}} \simeq 0.002 \, \text{fm}^2 $.
}
\begin{equation}
R_{c}^2
=
(R_{p}^{\text{point}})^2
+
\langle r_{p}^2 \rangle
+
\dfrac{N}{Z} \, \langle r_{n}^2 \rangle_{c}
,
\label{RcRp}
\end{equation}
where
$R_{p}^{\text{point}}$ is the point-proton distribution radius,
$
\langle r_{p}^2 \rangle^{1/2}
=
0.8414 \pm 0.0019  \, \text{fm} $~\cite{Hammer:2019uab}
is the charge radius of the proton
and
$ \langle r_{n}^2 \rangle_{c} = - 0.1161 \pm 0.0022 \, \text{fm}^2 $
is the squared charge radius of the neutron~\cite{Tanabashi:2018oca}.
Since the proton form factor
$F_{Z}(|\vet{q}|^2)$
in the cross section in Eq. (\ref{cs-std})
describes only the interaction of the protons in the nucleus,
the corresponding proton distribution radius $R_{p}$ is given by
\begin{equation}
R_{p}^2
=
(R_{p}^{\text{point}})^2
+
\langle r_{p}^2 \rangle
=
R_{c}^2
-
\dfrac{N}{Z} \, \langle r_{n}^2 \rangle_{c}
\,.
\label{Rp-th}
\end{equation}
From the experimental value of $R_{c}$ in Eq.~(\ref{Rc-exp}),
we obtain
\begin{equation}
R_{p} = 3.448 \pm 0.003 \, \text{fm}
\,.
\label{Rp}
\end{equation}
This is the value of the rms radius $R_{p}$ that we used in our calculations.

\begin{table}[!t]
\begin{center}
\begin{tabular}{rcc}
Interaction
&
$R_{p}^{\text{point}}$
&
$R_{n}^{\text{point}}$
\\
\hline
\multicolumn{3}{c}{Sky3D}
\\
SkI3 \protect\cite{Reinhard:1995zz}
&
3.33
&
3.43
\\
SkI4 \protect\cite{Reinhard:1995zz}
&
3.31
&
3.41
\\
Sly4 \protect\cite{Chabanat:1997un}
&
3.38
&
3.46
\\
Sly5 \protect\cite{Chabanat:1997un}
&
3.37
&
3.45
\\
Sly6 \protect\cite{Chabanat:1997un}
&
3.36
&
3.44
\\
Sly4d \protect\cite{Kim-Otsuka-Bonche-1997}
&
3.35
&
3.44
\\
SV-bas \protect\cite{Klupfel:2008af}
&
3.33
&
3.42
\\
UNEDF0 \protect\cite{Kortelainen:2010hv}
&
3.37
&
3.47
\\
UNEDF1 \protect\cite{Kortelainen:2011ft}
&
3.33
&
3.43
\\
SkM* \protect\cite{Bartel:1982ed}
&
3.37
&
3.45
\\
SkP \protect\cite{Dobaczewski:1983zc}
&
3.40
&
3.48
\\
\multicolumn{3}{c}{DIRHB}
\\
DD-ME2 \protect\cite{Niksic:2002yp}
&
3.30
&
3.39
\\
DD-PC1 \protect\cite{Niksic:2008vp}
&
3.30
&
3.39
\\
\hline
\end{tabular}
\end{center}
\caption{ \label{tab:models}
Values of the $^{40}\text{Ar}$
point-proton radius $R_{p}^{\text{point}}$
and
point-neutron radius $R_{n}^{\text{point}}$
obtained with the
Sky3D~\protect\cite{Maruhn:2013mpa}
and
DIRHB~\protect\cite{Niksic:2014dra}
codes with different nuclear interactions.
}
\end{table}

Let us now consider the neutron distribution radius
$R_{n}$ that determines the neutron form factor $F_{N}(|\vet{q}|^2)$
in the cross section in Eq.~(\ref{cs-std}).
Experimentally, the value of
$R_{n}$ is not known and we can get information on it
from the fit of the COHERENT data,
as discussed in Section~\ref{sec:neutron}.
However,
in our analysis it would be unphysical to consider $R_{n}$
as a completely free parameter,
because it is very plausible that
the neutron distribution radius is larger than the proton distribution radius
$R_{p}$ in Eq.~(\ref{Rp}),
since the $^{40}\text{Ar}$ nucleus has 22 neutrons and only 18 protons.
In order to check if this hypothesis is supported by the nuclear theory,
we have calculated the proton and neutron radii
with two publicly available numerical codes:
the Sky3D code~\cite{Maruhn:2013mpa}
of nonrelativistic nuclear mean-field models based on Skyrme forces,
and the
DIRHB code~\cite{Niksic:2014dra}
of relativistic self-consistent mean-field models.
Table~\ref{tab:models} presents the results of the calculation of the
point-proton radius $R_{p}^{\text{point}}$
and
point-neutron radius $R_{n}^{\text{point}}$
for different nuclear interactions
(the codes can calculate only the point-nucleon distributions,
that do not take into account the finite size of the nucleons).
From Table~\ref{tab:models} one can see that
$R_{n}^{\text{point}} > R_{p}^{\text{point}}$
in all the nuclear models that we have considered
and the excess is between 0.08 and 0.11 fm.
Since
\begin{equation}
R_{n}^2
=
(R_{n}^{\text{point}})^2
+
\langle r_{n}^2 \rangle
\,,
\label{Rn-th1}
\end{equation}
where
$ \langle r_{n}^2 \rangle^{1/2} \simeq \langle r_{p}^2 \rangle^{1/2} $
is the radius of the neutron
(this approximation is supported by
the measured value of the neutron magnetic radius
$
\langle r_{n}^2 \rangle_{\textrm{mag}}^{1/2}
=
0.864 {}^{+0.009}_{-0.008} \, \text{fm}
$~\cite{Tanabashi:2018oca},
that is close to the measured value of the proton charge radius
$
\langle r_{p}^2 \rangle^{1/2}
=
0.8414 \pm 0.0019  \, \text{fm}
$~\cite{Hammer:2019uab}).
Hence,
from the nuclear model prediction
$ R_{n}^{\text{point}} \simeq R_{p}^{\text{point}} + 0.1 \, \text{fm} $
we obtain the approximate relation
\begin{equation}
R_{n} \simeq R_{p} + 0.1 \, \text{fm}
\,.
\label{rn-th2}
\end{equation}
Therefore,
in our analyses of the COHERENT Argon data we consider two cases:
\begin{description}

\item[\textbf{Fixed $\mathbf{R_{n}}$}]
where $R_{n}$ is given by Eq.~(\ref{rn-th2})
with the value in Eq.~(\ref{Rp}) for $R_{p}$:
\begin{equation}
R_{n} = 3.55 \, \text{fm}
.
\label{Rnfixed}
\end{equation}

\item[\textbf{Free $\mathbf{R_{n}}$}]
where $R_{n}$ is considered as a free parameter
between $R_{p}$ and 4 fm:
\begin{equation}
3.45 < R_{n} < 4 \, \text{fm}
.
\label{Rnfree}
\end{equation}

\end{description}

The CE$\nu$NS event rate in the COHERENT experiment~\cite{Akimov:2020pdx}
depends on the neutrino flux
$d N_{\nu} / d E$ produced from the Spallation Neutron Source (SNS) at Oak Ridge Spallation Neutron Source.
It is given by the sum of
\begin{align}
\frac{d N_{\nu_{\mu}}}{d E}
=
\null & \null
\eta
\,
\delta\!\left(
E - \dfrac{ m_{\pi}^2 - m_{\mu}^2 }{ 2 m_{\pi} }
\right)
,
\label{numu}
\\
\frac{d N_{\nu_{\bar\mu}}}{d E}
=
\null & \null
\eta
\,
\dfrac{ 64 E^2 }{ m_{\mu}^3 }
\left(
\dfrac{3}{4} - \dfrac{E}{m_{\mu}}
\right)
,
\label{numubar}
\\
\frac{d N_{\nu_{e}}}{d E}
=
\null & \null
\eta
\,
\dfrac{ 192 E^2 }{ m_{\mu}^3 }
\left(
\dfrac{1}{2} - \dfrac{E}{m_{\mu}}
\right)
,
\label{nue}
\end{align}
with the normalization factor
$ \eta = r N_{\text{POT}} / 4 \pi L^2 $,
where
$r=(9\pm0.9)\times10^{-2}$ is the number of neutrinos per flavor
that are produced for each proton-on-target (POT),
$ N_{\text{POT}} = 13.7\times10^{22} $
is the number of proton on target corresponding to a total integrated beam power of 6.12 GW$\cdot$hr
and $ L = 27.5 \, \text{m} $
is the distance between the source and the COHERENT Ar detector, called CENNS-10~\cite{Akimov:2017ade}.
The pions decay at rest ($\pi^+\to \mu^++\nu_\mu$) producing $\nu_\mu$'s which arrive at the COHERENT detector as a prompt signal within about
$1.5 \, \mu\text{s}$
after protons-on-targets. The decay at rest of $\mu^{+}$ ($\mu^{+} \to e^{+} + \nu_{e} + \bar\nu_{\mu}$) produces
a delayed component of $\bar\nu_{\mu}$'s and $\nu_e$'s, since they arrive at the detector in a relatively longer time interval of about $10 \, \mu\text{s}$. 
In order to extract the physical parameter of interest, the first step is to simulate the \cenns signal at CENNS-10 as a function of the nuclear recoil energy. The theoretical \cenns event number $N_i^{\mathrm{CE\nu NS}}$ in each nuclear recoil energy bin $i$ is given by
\begin{equation}
\label{eq:en_spectra}
 N_i^{\mathrm{CE\nu NS}} = N(\mathrm{Ar}) \int^{T_{\mathrm{nr}}^{i+1}}_{T_{\mathrm{nr}}^{i}} dT_{\mathrm{nr}}\,A(T_{\mathrm{nr}}) \int^{E_{\mathrm{max}}}_{E_{\mathrm{min}}} dE \sum_{\nu=\nu_e,\nu_\mu,\overline{\nu}_\mu} \frac{dN_\nu}{dE}  \dfrac{d\sigma_{\nu\text{-}\mathcal{N}}}{d T_\mathrm{nr}}(E,T_\mathrm{nr})\,,
\end{equation}
where $A(T_{\mathrm{nr}})$ is the energy-dependent  reconstruction efficiency given in Fig.~3 in Ref.~\cite{Akimov:2020pdx}, $E_{\mathrm{min}} = \sqrt{MT_\mathrm{nr}/2}$ and $E_{\mathrm{max}} = m_\mu/2 \sim 52.8$ MeV, $m_\mu$ being the muon mass, $N(\mathrm{Ar})$ is the number of Ar atoms in the detector, and $\frac{dN_\nu}{dE}$ is the neutrino 
flux integrated over the experiment lifetime. Concerning the former element, we digitalise the efficiency as a function of the electron-equivalent recoil energy $T_{ee}\,[\mathrm{keV}_{ee}]$, which is subsequently transformed as a function of the nuclear recoil energy $T_{\mathrm{nr}}\,[\mathrm{keV}_{\mathrm{nr}}]$ thanks to the relation 
\begin{equation}
\label{eq:qf}
T_{ee} = f_Q (T_{\mathrm{nr}})T_{\mathrm{nr}}\,.
\end{equation}
Here, $f_Q$ is the quenching factor, which is the ratio between the scintillation light emitted in nuclear and electron recoils
and determines the relation between the number of detected photoelectrons and the nuclear recoil kinetic energy. Following Ref.~\cite{Akimov:2020pdx}, the quenching factor is parameterized as $f_Q (T_{\mathrm{nr}}) = (0.246\pm0.006\, \mathrm{keV}_{\mathrm{nr}}) + ((7.8\pm0.9)\times10^{-4})T_{\mathrm{nr}}$ up to $125\,\mathrm{keV}_{\mathrm{nr}}$, and kept constant for larger values. The value of $N(\mathrm{Ar})$ is given by $N_\mathrm{A}\,M_{\mathrm{det}}/M_{\mathrm{Ar}}$, where $N_\mathrm{A}$ is the Avogadro number, $M_{\mathrm{det}}$ is the detector active mass equal to 24 kg and $M_{\mathrm{Ar}}=39.96$ g/mol is the molar mass of $^{40}\mathrm{Ar}$. Actually, one should consider that atmospheric argon is contaminated by a small percentage of $^{36}\mathrm{Ar}$ and $^{38}\mathrm{Ar}$, namely $F( ^{36}\mathrm{Ar})=0.334\%$ and $F( ^{38}\mathrm{Ar})=0.063\%$. However, since the amount of $^{36}\mathrm{Ar}$ and $^{38}\mathrm{Ar}$ is very small
and the uncertainties are large, in practice one gets the same results considering
$F( ^{40}\mathrm{Ar})=100\%$ and $F( ^{36}\mathrm{Ar})=F( ^{38}\mathrm{Ar})=0$.

In Ref.~\cite{Akimov:2020pdx} two independent analyses, labeled A and B, are described, that differ mainly for the selection and the treatment of the background. In the following, we will use the data coming from the analysis A, whose range of interest of the nuclear recoil energy is [0, 120]~$\mathrm{keV}_{ee}$ (corresponding to roughly [0, 350]~$\mathrm{keV}_{\mathrm{nr}}$), with 12 energy bins of size equal to 10~$\mathrm{keV}_{ee}$. 
We have also performed the analyses of the data corresponding to analysis B
of the COHERENT collaboration~\cite{Akimov:2020pdx}
described in appendix~\ref{app:analysisB},
where we considered only
the determination of the radius of the nuclear neutron distribution
and of the weak mixing angle. 

In our analysis corresponding to analysis A
of the COHERENT collaboration~\cite{Akimov:2020pdx},
we considered the least-squares function
\begin{eqnarray}
\chi^2_{\text{S}}
&=&
\sum_{i=1}^{12}
\left(
\dfrac{
N_{i}^{\text{exp}}
-
\eta_{\mathrm{CE\nu NS}} N_i^{\mathrm{CE\nu NS}}
-
\eta_{\mathrm{PBRN}} B_i^{\mathrm{PBRN}}
-
\eta_{\mathrm{LBRN}} B_i^{\mathrm{LBRN}}}
{\sigma_i}
\right)^2\\ \nonumber
&+&
\left( \dfrac{\eta_{\mathrm{CE\nu NS}}-1}{\sigma_{\mathrm{CE\nu NS}}} \right)^2
+
\left( \dfrac{\eta_{\mathrm{PBRN}}-1}{\sigma_{\mathrm{PBRN}}} \right)^2
+
\left( \dfrac{\eta_{\mathrm{LBRN}}-1}{\sigma_{\mathrm{LBRN}}} \right)^2
,
\label{chi-spectrum}
\end{eqnarray}
where PBRN stands for Prompt Beam-Related Background, LBRN for Late Beam-Related Neutron Background and with
\begin{eqnarray}
\sigma_i^2 = \left( \sigma_i^{\mathrm{exp}} \right)^2 &+&
\left[ \sigma_{\mathrm{BRNES}} \left( B_i^{\mathrm{PBRN}} + B_i^{\mathrm{LBRN}}\right)\right]^2,\\
\sigma_{\mathrm{BRNES}} &=& \sqrt{\frac{0.058^2}{12}}=1.7\%,\\
\sigma_{\mathrm{CE\nu NS}} &=& 13.4\%\,\mathrm{for\,fixed\,}R_{n},\,\mathrm{or}\,13.2\%\,\mathrm{for\,free\,}R_{n},\\
\sigma_{\mathrm{PBRN}} &=& 32\%,\\
\sigma_{\mathrm{LBRN}} &=& 100\%.
\end{eqnarray}
For each energy bin $i$,
$N_{i}^{\text{exp}}$ is the experimental event number,
$N_{i}^{\mathrm{CE\nu NS}}$
is the theoretical event number
that is calculated as explained in Section~\ref{sec:theory},
$B_i^{\mathrm{PBRN}}$ and $B_i^{\mathrm{LBRN}}$ are the estimated number of PBRN and LBRN background events, and
$\sigma_{i}$ is the total signal uncertainty.
The Beam Related Neutron Energy Shape (BRNES) 5.8$\%$ uncertainty ($\sigma_{\mathrm{BRNES}}$) is taken into account by distributing it over the 12 bins in an uncorrelated way. All the numbers are taken from Ref.~\cite{Akimov:2020pdx}.

In Eq.~(\ref{chi-spectrum}),
$\eta_{\mathrm{CE\nu NS}}$, $\eta_{\mathrm{PBRN}}$ and $\eta_{\mathrm{LBRN}}$ 
are nuisance parameters which quantify,
respectively,
the systematic uncertainty of the signal rate
and
the systematic uncertainty of the PBRN and LBRN background rate,
with
corresponding standard deviations
$\sigma_{\mathrm{CE\nu NS}}$, $\sigma_{\mathrm{PBRN}}$
and
$\sigma_{\mathrm{LBRN}}$.

\begin{figure*}[!t]
\centering
\setlength{\tabcolsep}{0pt}
\begin{tabular}{cc}
\subfigure[]{\label{fig:hist1}
\begin{tabular}{c}
\includegraphics*[width=0.49\linewidth]{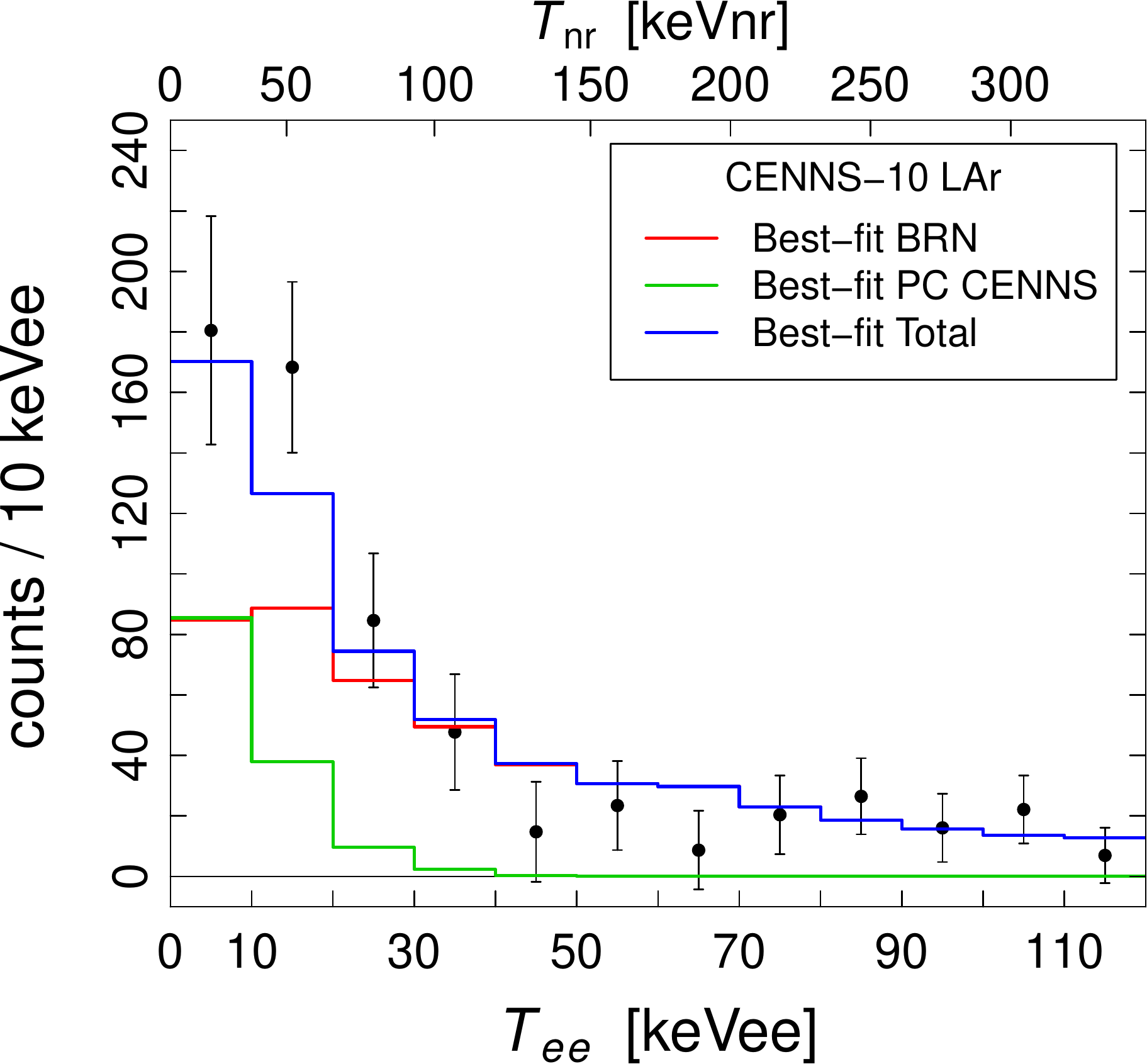}
\\
\end{tabular}
}
&
\subfigure[]{\label{fig:hist2}
\begin{tabular}{c}
\includegraphics*[width=0.49\linewidth]{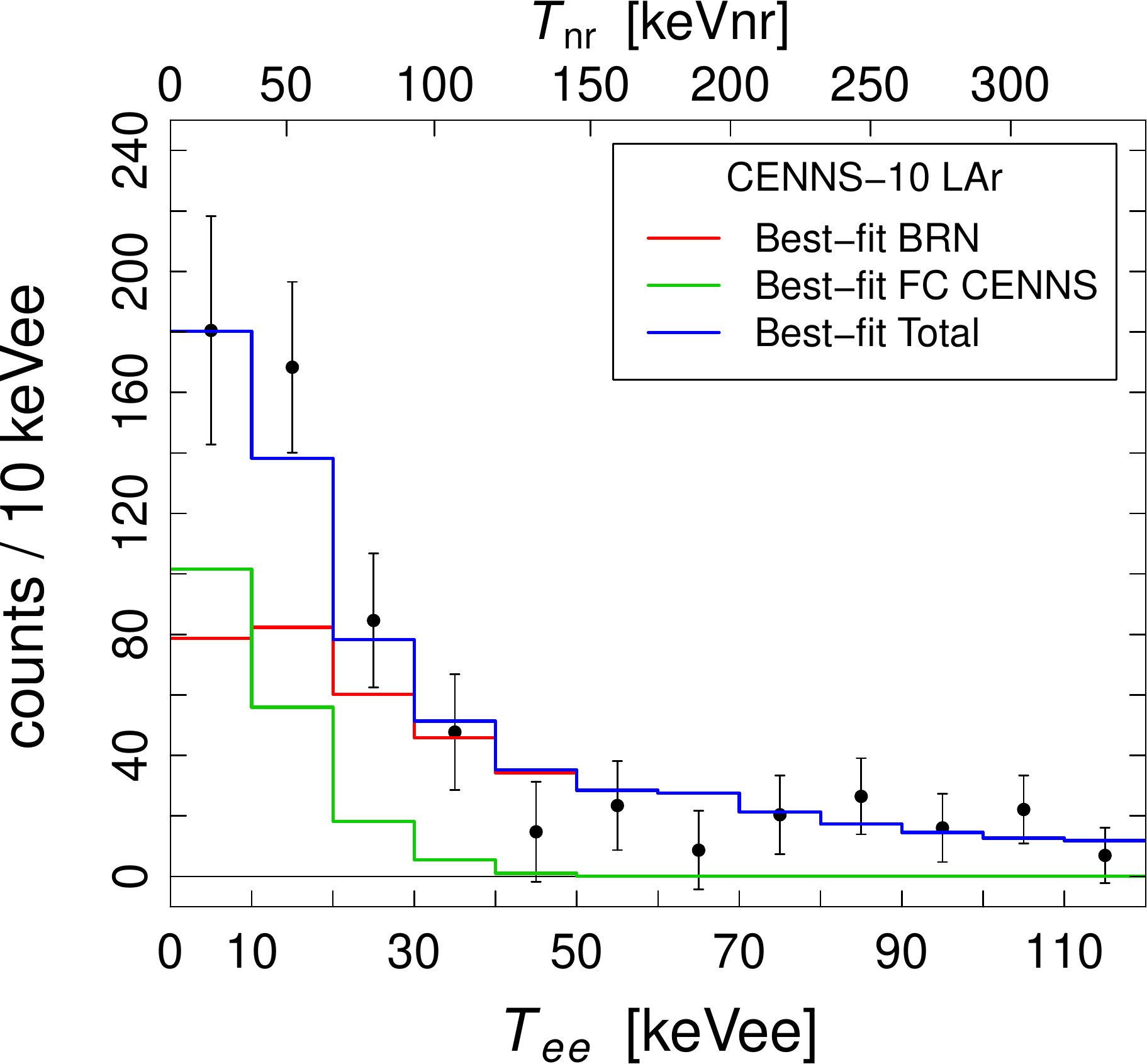}
\\
\end{tabular}
}
\end{tabular}
\caption{ \label{fig:hist}
Histograms representing the fits of the CENNS-10 data
(black points with statistical error bars)
in the case of
\subref{fig:hist1} partial coherence (PC),
with the neutron form factor
corresponding to the minimal neutron distribution radius
$ R_{n} = 3.45 \, \text{fm}$,
and
\subref{fig:hist2} full coherence (FC).
}
\end{figure*}

The COHERENT spectral data are shown in Figure~\ref{fig:hist1}
together with the best-fit histogram obtained with
the CE$\nu$NS cross section of Eq.~(\ref{cs-std})
and the neutron form factor
corresponding to the minimal neutron distribution radius
$ R_{n} = 3.45 \, \text{fm}$,
that gives the larger CE$\nu$NS cross section
for $R_{n}$ in the range in Eq.~(\ref{Rnfree}).
We obtained
$ (\chi^2_{\text{S}})_{\text{min}} = 8.8 $
with
$11$
degrees of freedom,
corresponding to an excellent
$64$\%
goodness of fit.
We tested also the case of full coherence, i.e. without the suppression of the neutron and proton form factors, as done in the case of the COHERENT CsI
data~\cite{Cadeddu:2017etk,Cadeddu:2019eta}.
In this case, illustrated in Figure~\ref{fig:hist2},
the larger CE$\nu$NS cross section fits slightly better the
low-energy data and the medium- and high-energy data
are fitted slightly better with a smaller background within the uncertainties.
Indeed, the full coherence $(\chi^2_{\text{S}})_{\text{min}}$ is
$7.0$,
that is smaller than the
$8.8$
obtained with the minimal neutron distribution radius.
However,
we will not consider the full coherence in the rest of the paper,
because we are not aware of any physical mechanism that
can justify the absence of the form-factor suppression
corresponding to the physical nucleon distributions in the nucleus.

\section{Radius of the nuclear neutron distribution}
\label{sec:neutron}

The observation of \cenns scattering in argon
can be used to probe the nuclear neutron distribution~\cite{Patton:2012jr,Cadeddu:2017etk,Papoulias:2019lfi,Ciuffoli:2018qem,Papoulias:2019lfi}.
We fitted the COHERENT data in order to determine the neutron rms radius
$R_{n}$
of Ar,
considering
for $R_{n}$ the lower bound in Eq.~(\ref{Rnfree}),
without an upper bound.

\begin{figure}[!t]
\centering
\includegraphics*[width=0.5\linewidth]{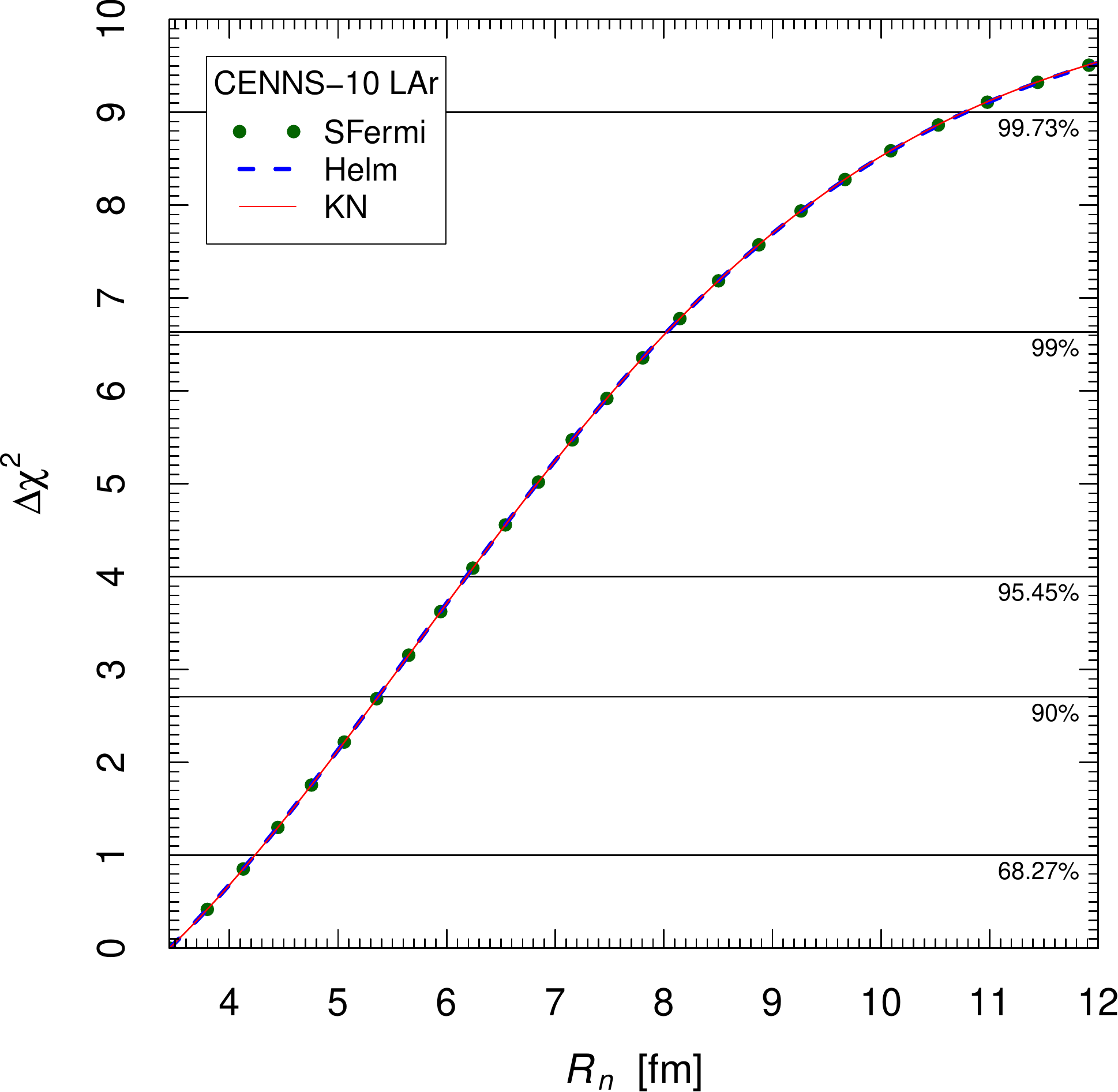}
\caption{ \label{fig:rn-chi}
$\Delta\chi^2 = \chi^2_{\text{S}} - (\chi^2_{\text{S}})_{\text{min}}$
as a function of the rms neutron distribution radius $R_{n}$
of $^{40}\text{Ar}$
obtained from the fit of the data of the CENNS-10 experiment.
The three curves correspond to the
symmetrized Fermi~\cite{Piekarewicz:2016vbn} (SFermi),
Helm~\cite{Helm:1956zz} (Helm), and
Klein-Nystrand~\cite{Klein:1999qj} (KN)
form factor parameterizations.
}
\end{figure}

Figure~\ref{fig:rn-chi} shows the comparison of
$\Delta\chi^2 = \chi^2_{\text{S}} - (\chi^2_{\text{S}})_{\text{min}}$
as a function of the rms neutron distribution radius $R_{n}$ of $^{40}\text{Ar}$
using the three most popular form factor parameterizations:
symmetrized Fermi~\cite{Piekarewicz:2016vbn},
Helm~\cite{Helm:1956zz}, and
Klein-Nystrand~\cite{Klein:1999qj}.
One can see that the three form factor parameterizations
give practically the same result
and the best fit is obtained for the minimal allowed value
$R_{n} = 3.45 \, \text{fm}$.
Therefore,
from the analysis of the COHERENT data
we can only put the following upper bounds on the value of
$^{40}\text{Ar}$ neutron distribution radius:
\begin{equation}
R_{n}({}^{40}\text{Ar})
<
4.2
\, (1\sigma)
,
\,
6.2
\, (2\sigma)
,
\,
10.8
\, (3\sigma)
\, \text{fm}
.
\label{Rn-bounds}
\end{equation}
These bounds are in agreement with the nuclear model predictions
in Table~\ref{tab:models},
but unfortunately they are too weak to allow us a selection of the models.

\section{Weak mixing angle}
\label{sec:electroweak}

\begin{figure}[!t]
\centering
\includegraphics*[width=0.5\linewidth]{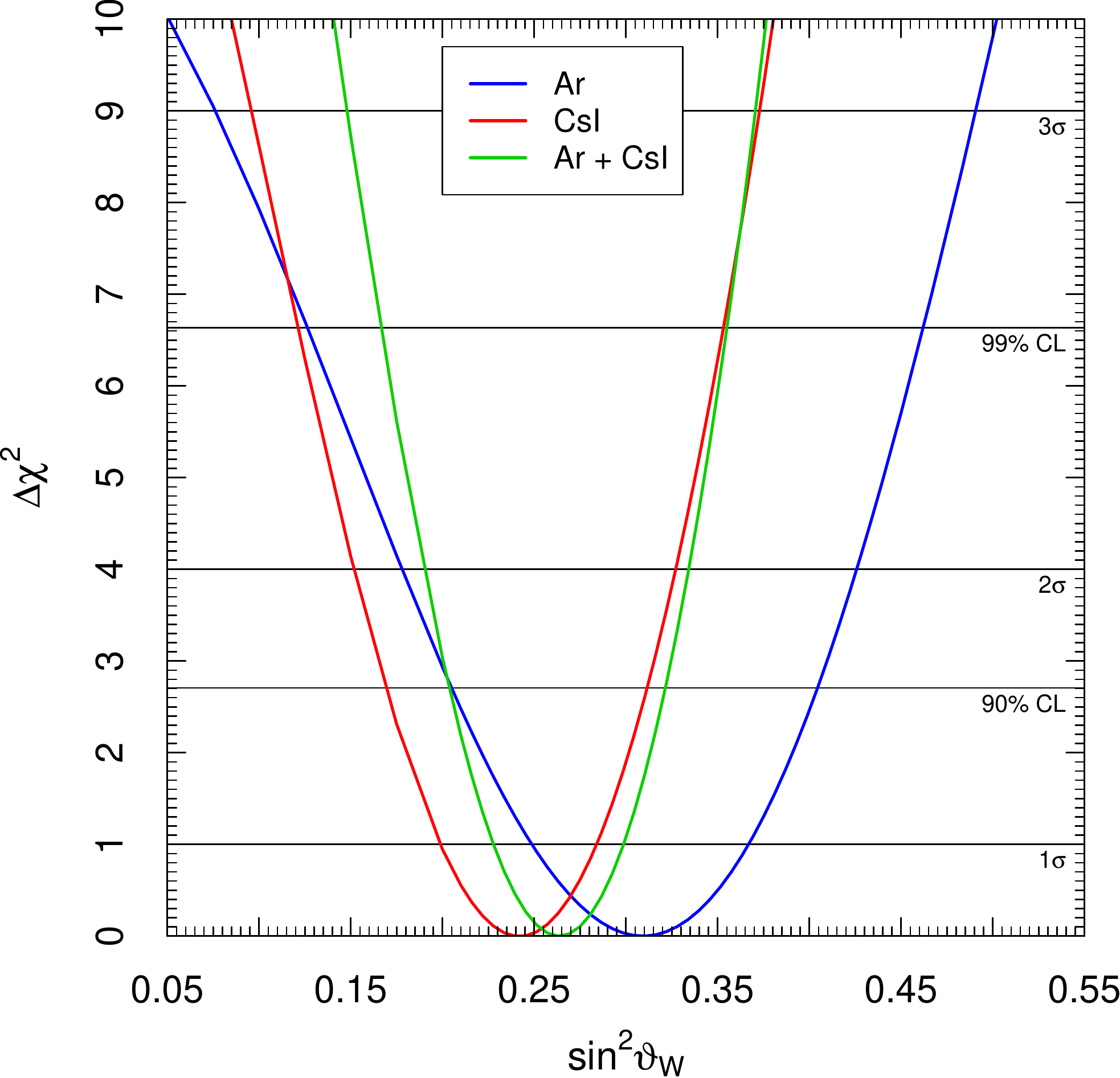}
\caption{ \label{fig:thetaw-chi}
$\Delta\chi^2 = \chi^2_{\text{S}} - (\chi^2_{\text{S}})_{\text{min}}$
as a function $\sin^2{\vartheta_{\text{W}}}$ 
obtained (blue) from the fit of the data of the Ar CENNS-10 experiment, (red) from the fit of the COHERENT CsI data and (green) from the combined fit .
}
\end{figure}

The weak mixing angle is a fundamental parameter in the theory of the EW interactions and its experimental determination provides a direct  probe  of  physics  phenomena  not  included  in  the SM, usually referred to as new physics. In particular, low-energy determinations of $\vartheta_{\text{W}}$ offer a unique role, complementary to those at high-energy, being highly sensitive to extra $Z$ ($Z'$) bosons predicted  in grand unified theories,  technicolor models, supersymmetry and string theories~\cite{Safronova_2018}. This underscores  the need  for improved  experimental determinations of $\vartheta_{\text{W}}$ in the low-energy regime. \\
We fitted the COHERENT CENNS-10 data in order to determine the value of
$\sin^2{\vartheta_{\text{W}}}$ in Ar,
considering
$R_{n}$ either fixed or free. 
The result for the weak mixing angle is independent on the assumption used for $R_{n}$ and in both cases we get: 
\begin{equation}
  \sin^2{\vartheta_{\text{W}}}(\mathrm{Ar}) = 0.31\pm 0.06\,(1\sigma),
  ^{+0.11}_{-0.13}\,(2\sigma),
  ^{+0.18}_{-0.23}\,(3\sigma),
\end{equation}
which is about 1.2$\sigma$ above the SM prediction, $\sin^2{\vartheta_{\text{W}}^{\mathrm{SM}}}=0.23857(5)$~\cite{Tanabashi:2018oca}. The reason of this small discrepancy is that a larger weak mixing angle increases the \cenns cross section and it allows a better fit of the low-energy bins of the Ar data.
Given the independence of $\sin^2{\vartheta_{\text{W}}}$ on the value of $R_{n}$, in the following we will consider only the case with $R_{n}$ fixed.
Figure~\ref{fig:thetaw-chi} shows the comparison of
$\Delta\chi^2 = \chi^2_{\text{S}} - (\chi^2_{\text{S}})_{\text{min}}$
as a function of $\sin^2{\vartheta_{\text{W}}}$
using the Helm parameterization for the neutron form factor.

\begin{figure}[!t]
\centering
\includegraphics*[width=0.7\linewidth]{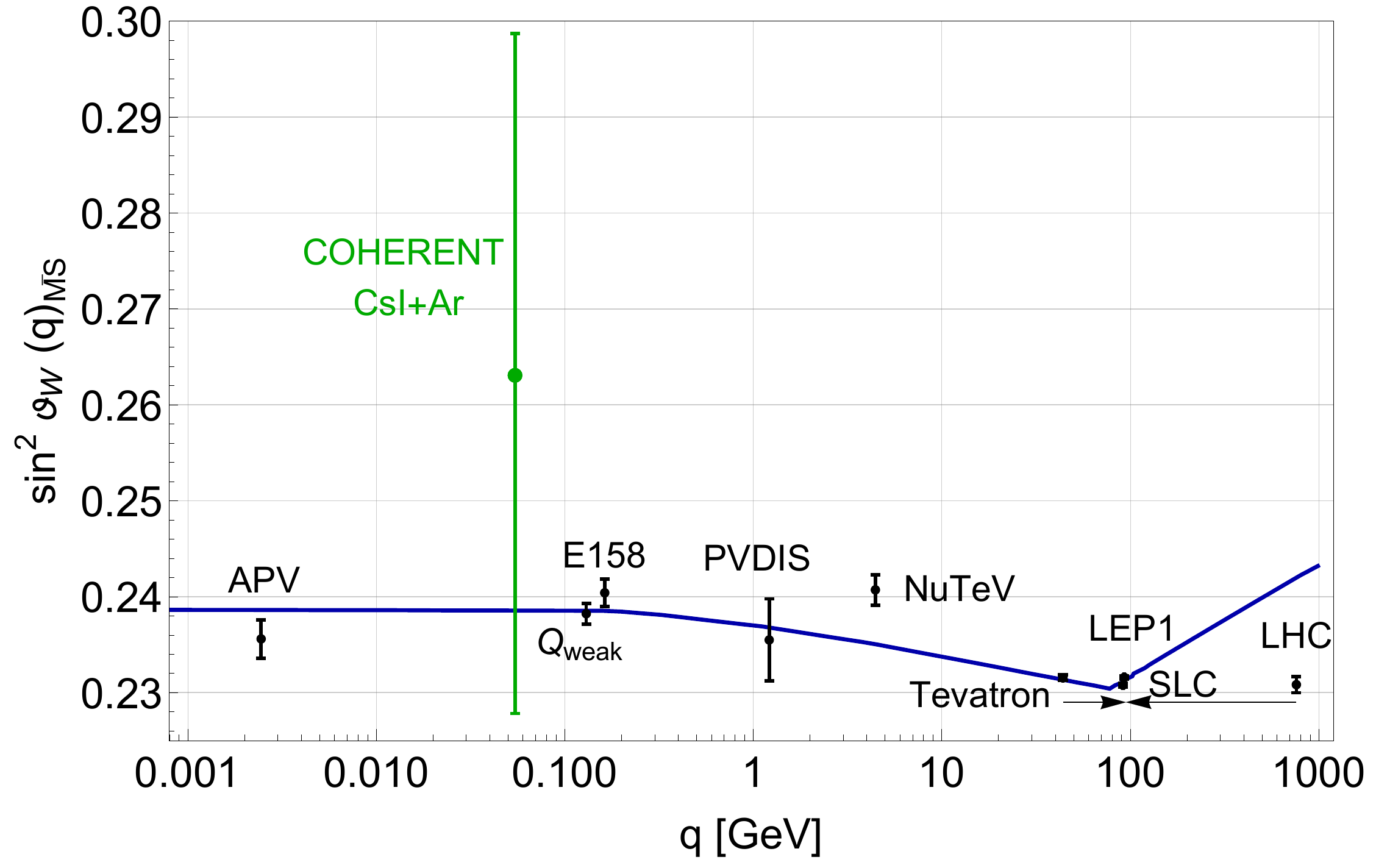}
\caption{ \label{fig:running}
Variation of $\sin^2 \vartheta_{\text{W}}$ with energy scale q. The SM prediction is shown as the solid curve, together with
experimental determinations in black at the $Z$-pole~\cite{Tanabashi:2018oca} (Tevatron, LEP1, SLC, LHC),
from APV on cesium~\cite{Wood:1997zq,Dzuba:2012kx}, which has a typical momentum transfer given by $\langle q\rangle\simeq$~2.4 MeV, M{\o}ller scattering~\cite{Anthony:2005pm} (E158), deep inelastic scattering of polarized electrons on deuterons~\cite{Wang:2014bba} ($ e^2H $ PVDIS) and from
neutrino-nucleus scattering~\cite{Zeller:2001hh} (NuTeV) and the new result from the proton's weak charge 
at $q = 0.158$ GeV~\cite{Androic:2018kni} ($ \text{Q}_{\text{weak}} $). In green it is shown the result derived in this paper, obtained fitting the Ar and CsI COHERENT dataset. For clarity we displayed the Tevatron and LHC points horizontally to the left and to the right, respectively.}
\end{figure}

Following the approach used in Ref.~\cite{Cadeddu:2019eta}, where we improved the bounds on several physical quantities from the analysis of the COHERENT CsI data~\cite{Akimov:2017ade} 
considering the improved quenching factor in Ref.~\cite{Collar:2019ihs}, we derive here the result for the weak mixing angle also exploiting the COHERENT CsI dataset. Fixing $R_n(\mathrm{Cs})$ and $R_n(\mathrm{I})$ to 5.01~fm and 4.94~fm~\cite{Bender_1999}, respectively, we get
\begin{equation}
\sin^2{\vartheta_{\text{W}}}(\mathrm{CsI})= 0.24\pm0.04\,(1\sigma),
  \pm 0.09\,(2\sigma),
  ^{+0.13}_{-0.14}\,(3\sigma),
\end{equation}
in very good agreement with the SM prediction. The corresponding $\Delta\chi^2$ is also shown in Figure~\ref{fig:thetaw-chi}.\\
Finally, we performed a combined fit of the CsI and Ar data. The value found for the weak mixing angle is
\begin{equation}
\sin^2{\vartheta_{\text{W}}}(\mathrm{CsI+Ar})= 0.26^{+0.04}_{-0.03}\,(1\sigma),  \pm 0.07\,(2\sigma), \pm 0.11\,(3\sigma),
\end{equation}
which is slightly more precise than the CsI result alone and in agreement within 1$\sigma$ with the SM prediction. Unfortunately, as it is possible to see in 
Figure~\ref{fig:running}, the uncertainty obtained for the weak mixing angle from COHERENT is still very large when compared to the other determinations at low-momentum transfer. 

For the proton coupling coefficient $g_{V}^{p}$, we obtain
\begin{equation}
g_{V}^{p}(\nu_{e};\mathrm{CsI+Ar})
=
-0.003
{}^{+0.060}_{-0.080}
\quad
\text{and}
\quad
g_{V}^{p}(\nu_{\mu};\mathrm{CsI+Ar})
=
-0.011
{}^{+0.060}_{-0.080}
.
\label{gVpCsIAr}
\end{equation}
These values differ from the SM predictions in Eqs.~(\ref{gVp-nue}) and (\ref{gVp-num})
by less than $1\sigma$ and confirm that the proton coupling
is much smaller than the neutron coupling in the \cenns process.

\section{Neutrino charge radii}
\label{sec:radii}

The neutrino charge radii are the only electromagnetic properties
of neutrinos that are nonzero in the Standard Model of electroweak interactions.
They are induced by radiative corrections,
with the predicted values
\cite{Bernabeu:2000hf,Bernabeu:2002nw,Bernabeu:2002pd}
\begin{equation}
\langle{r}_{\nu_{\ell}}^2\rangle_{\text{SM}}
=
-
\frac{G_{\text{F}}}{2\sqrt{2}\pi^2}
\left[
3-2\ln\left(\frac{m_{\ell}^2}{m^2_{W}}\right)
\right]
,
\label{G050}
\end{equation}
where $m_{W}$ and $m_{\ell}$ are the $W$ boson and charged lepton masses
($\ell = e, \mu, \tau$),
and we use the conventions in Refs.~\cite{Giunti:2014ixa,Cadeddu:2018dux,Cadeddu:2019eta}.
The Standard Model charge radii of neutrinos are diagonal in the flavor basis,
because in the Standard Model the generation lepton numbers are conserved.
Numerically, the predicted values of
$\langle{r}_{\nu_{e}}^2\rangle_{\text{SM}}$
and
$\langle{r}_{\nu_{\mu}}^2\rangle_{\text{SM}}$,
that can be probed with the data of the COHERENT experiment,
are
\begin{align}
\null & \null
\langle{r}_{\nu_{e}}^2\rangle_{\text{SM}}
=
- 0.83 \times 10^{-32} \, \text{cm}^2
,
\label{reSM}
\\
\null & \null
\langle{r}_{\nu_{\mu}}^2\rangle_{\text{SM}}
=
- 0.48 \times 10^{-32} \, \text{cm}^2
.
\label{rmSM}
\end{align}
The current 90\% CL experimental bounds for
$\langle{r}_{\nu_{e}}^2\rangle$
and
$\langle{r}_{\nu_{\mu}}^2\rangle$
obtained in laboratory experiments that do not involve
CE$\nu$NS are listed in Table~I of Ref.~\cite{Cadeddu:2018dux}.
Since they are
only about one order of magnitude larger than the Standard Model predictions,
they may be the first neutrino electromagnetic properties
measured by new experiments in a near future.

As discussed in Section~\ref{sec:theory}
the contribution of the Standard Model charge radius of $\nu_{\ell}$
is taken into account by the last term in the expression (\ref{gVpth})
of $g_{V}^{p}(\nu_{\ell})$.
Here, we want to study the effects of the neutrino charge radii in the
CE$\nu$NS data of the COHERENT experiment
independently of the origin of the charge radii,
that can have contributions both from the Standard Model
and from physics beyond the Standard Model.
Therefore we consider the differential cross section
\begin{equation}
\dfrac{d\sigma_{\nu_{\ell}\text{-}\mathcal{N}}}{d T_\mathrm{nr}}
(E,T_\mathrm{nr})
=
\dfrac{G_{\text{F}}^2 M}{\pi}
\left(
1 - \dfrac{M T_\mathrm{nr}}{2 E^2}
\right)
\left\{
\left[
\left( \tilde{g}_{V}^{p} - \tilde{Q}_{\ell\ell} \right)
Z
F_{Z}(|\vet{q}|^2)
+
g_{V}^{n}
N
F_{N}(|\vet{q}|^2)
\right]^2
+
Z^2
F_{Z}^2(|\vet{q}|^2)
\sum_{\ell'\neq\ell}
|\tilde{Q}_{\ell'\ell}|^2
\right\}
,
\label{cs-chr}
\end{equation}
where $\tilde{g}_{V}^{p}=0.0204$
is given by Eq.~(\ref{gVpth})
without the last term that contains the contribution of the Standard Model charge radius.
The effects of the charge radii
$\langle{r}_{\nu_{\ell\ell'}}^2\rangle$
in the cross section are expressed through~\cite{Kouzakov:2017hbc}
\begin{equation}
\tilde{Q}_{\ell\ell'}
=
\dfrac{ \sqrt{2} \pi \alpha }{ 3 G_{\text{F}} }
\, \langle{r}_{\nu_{\ell\ell'}}^2\rangle
.
\label{Qchr}
\end{equation}
We consider the general case in which neutrinos can have both diagonal and
off-diagonal charge radii in the flavor basis.
The off-diagonal charge radii,
as well as part of the diagonal charge radii,
can be generated by physics beyond the Standard Model.

The effects of the charge radii in the cross section are sometimes expressed
through~\cite{Grau:1985cn,Degrassi:1989ip}
\begin{equation}
\hat{Q}_{\ell\ell'}
=
\frac{2}{3} \, m_{W}^2 \sin^2\!\vartheta_{W} \langle{r}_{\nu_{\ell\ell'}}^2\rangle
,
\label{hatQchr}
\end{equation}
that is considered equivalent to $\tilde{Q}_{\ell\ell'}$ in Eq.~(\ref{Qchr})~\cite{Vogel:1989iv}
through the well-known relations
$ G_{\text{F}} / \sqrt{2} = g^2 / 8 m_{W}^2 $
and
$ g^2 \sin^2\!\vartheta_{W} = e^2 = 4 \pi \alpha $,
where $g$ is the weak charged-current coupling constant
and $e$ is the elementary electric charge
(see, for example, Ref.~\cite{Giunti:2007ry}).
The problem is that the equivalence holds only at tree level
and radiative corrections induce a significant difference.
Indeed, using the PDG values of all quantities~\cite{Tanabashi:2018oca}
we obtain, neglecting the uncertainties,
$
\sqrt{2} \pi \alpha / 3 G_{\text{F}}
=
2.38 \times 10^{30} \, \text{cm}^{-2}
$
and
$
2 m_{W}^2 \sin^2\!\vartheta_{W} / 3
=
2.64 \times 10^{30} \, \text{cm}^{-2}
$,
that differ by about 10\%.
Therefore,
the form in Eq. (\ref{hatQchr}) overestimates the effect of the charge radius
by about 10\%
with respect to the form in Eq. (\ref{Qchr}),
that is the correct one for low-energy interactions
because it depends only on measured low-energy quantities.
Moreover,
one can notice that the electromagnetic interaction due to the charge radius
must be proportional to the electromagnetic fine-structure constant $\alpha$
and must be independent of the Fermi weak interaction constant $G_{\text{F}}$.
Indeed, the $G_{\text{F}}$ in the denominator of Eq.~(\ref{Qchr})
cancels the $G_{\text{F}}$ in the cross section (\ref{cs-chr}).

The diagonal charge radii of  flavor neutrinos
contribute to the cross section coherently with the
neutrino-proton neutral current interaction,
generating an effective shift of $\sin^2\!\vartheta_{W}$.
In the case of
$\bar\nu_{\ell}\text{-}\mathcal{N}$ scattering,
we have
$g_{V}^{p,n} \to - g_{V}^{p,n}$
and
$\langle{r}_{\nu_{\ell\ell'}}\rangle
\to
\langle{r}_{\bar\nu_{\ell\ell'}}\rangle = - \langle{r}_{\nu_{\ell\ell'}}\rangle$.
Therefore,
the charge radii of flavor neutrinos and antineutrinos
contribute with the same sign to the shift of
$\sin^2\!\vartheta_{W}$
in the CE$\nu$NS cross section.

There are five charge radii that can be determined with the COHERENT
CE$\nu$NS data:
the two diagonal charge radii
$\langle{r}_{\nu_{ee}}^2\rangle$
and
$\langle{r}_{\nu_{\mu\mu}}^2\rangle$,
that sometimes are denoted with the simpler notation
$\langle{r}_{\nu_{e}}^2\rangle$
and
$\langle{r}_{\nu_{\mu}}^2\rangle$
in connection to the Standard Model charge radii in Eqs.~(\ref{G050})--(\ref{rmSM}),
and the absolute values of the three off-diagonal charge radii
$\langle{r}_{\nu_{e\mu}}^2\rangle=\langle{r}_{\nu_{\mu e}}^2\rangle^{*}$,
$\langle{r}_{\nu_{e\tau}}^2\rangle$, and
$\langle{r}_{\nu_{\mu\tau}}^2\rangle$.

In Ref.~\cite{Cadeddu:2018dux}
we obtained the bounds on the neutrino charge radii
from the analysis of the COHERENT CsI data~\cite{Akimov:2017ade}.
In Ref.~\cite{Cadeddu:2019eta} we improved these bounds
considering the improved quenching factor in Ref.~\cite{Collar:2019ihs}.
Here we present the bounds on the neutrino charge radii
that we obtained from the analysis of the
spectral Ar data of the COHERENT experiment~\cite{Akimov:2020pdx}
and those obtained with a combined fit of the CsI and Ar data.
We also revise the CsI limits on the charge radii presented
in Ref.~\cite{Cadeddu:2019eta}
because they have been obtained through Eq.~(\ref{hatQchr}),
that overestimates their contribution by about 10\%, as discussed above.

\begin{table*}[t!]
\begin{center}
\begin{tabular}{ccccccc}
\\
&
\multicolumn{3}{c}{Fixed $R_{n}$}
&
\multicolumn{3}{c}{Free $R_{n}$}
\\
&
$1\sigma$
&
$2\sigma$
&
$3\sigma$
&
$1\sigma$
&
$2\sigma$
&
$3\sigma$
\\
\hline
&
\multicolumn{6}{c}{CsI}
\\
$\langle{r}_{\nu_{ee}}^2\rangle$
&
$ -55 \div -2 $
&
$ -67 \div 11 $
&
$ -76 \div 20 $
&
$ -54 \div 1 $
&
$ -66 \div 14 $
&
$ -76 \div 24 $
\\
$\langle{r}_{\nu_{\mu\mu}}^2\rangle$
&
$ -64 \div 8 $
&
$ -68 \div 12 $
&
$ -73 \div 17 $
&
$ -64 \div 10 $
&
$ -68 \div 15 $
&
$ -72 \div 20 $
\\
$\langle{r}_{\nu_{e\mu}}^2\rangle$
&
$ < 26 $
&
$ < 32 $
&
$ < 37 $
&
$ < 26 $
&
$ < 32 $
&
$ < 36 $
\\
$\langle{r}_{\nu_{e\tau}}^2\rangle$
&
$ < 27 $
&
$ < 39 $
&
$ < 48 $
&
$ < 27 $
&
$ < 39 $
&
$ < 48 $
\\
$\langle{r}_{\nu_{\mu\tau}}^2\rangle$
&
$ < 36 $
&
$ < 40 $
&
$ < 45 $
&
$ < 36 $
&
$ < 40 $
&
$ < 45 $
\\
&
\multicolumn{6}{c}{Ar}
\\
$\langle{r}_{\nu_{ee}}^2\rangle$
&
$ -89 \div 39 $
&
$ -98 \div 48 $
&
$ -108 \div 58 $
&
$ -89 \div 38 $
&
$ -97 \div 47 $
&
$ -107 \div 57 $
\\
$\langle{r}_{\nu_{\mu\mu}}^2\rangle$
&
$ -63 \div 12 $
&
$ -73 \div 22 $
&
$ -80 \div 30 $
&
$ -63 \div 9 $
&
$ -72 \div 22 $
&
$ -80 \div 29 $
\\
$\langle{r}_{\nu_{e\mu}}^2\rangle$
&
$ < 34 $
&
$ < 40 $
&
$ < 46 $
&
$ < 33 $
&
$ < 40 $
&
$ < 46 $
\\
$\langle{r}_{\nu_{e\tau}}^2\rangle$
&
$ < 64 $
&
$ < 73 $
&
$ < 83 $
&
$ < 63 $
&
$ < 72 $
&
$ < 82 $
\\
$\langle{r}_{\nu_{\mu\tau}}^2\rangle$
&
$ < 37 $
&
$ < 48 $
&
$ < 55 $
&
$ < 36 $
&
$ < 47 $
&
$ < 54 $
\\
&
\multicolumn{6}{c}{CsI + Ar}
\\
$\langle{r}_{\nu_{ee}}^2\rangle$
&
$ -56 \div -2 $
&
$ -68 \div 11 $
&
$ -78 \div 22 $
&
$ -55 \div -4 $
&
$ -67 \div 14 $
&
$ -77 \div 25 $
\\
$\langle{r}_{\nu_{\mu\mu}}^2\rangle$
&
$ -64 \div 6 $
&
$ -68 \div 12 $
&
$ -71 \div 17 $
&
$ -64 \div 9 $
&
$ -67 \div 15 $
&
$ -71 \div 19 $
\\
$\langle{r}_{\nu_{e\mu}}^2\rangle$
&
$ < 27 $
&
$ < 33 $
&
$ < 36 $
&
$ < 25 $
&
$ < 32 $
&
$ < 36 $
\\
$\langle{r}_{\nu_{e\tau}}^2\rangle$
&
$ < 27 $
&
$ < 40 $
&
$ < 50 $
&
$ < 26 $
&
$ < 40 $
&
$ < 50 $
\\
$\langle{r}_{\nu_{\mu\tau}}^2\rangle$
&
$ < 36 $
&
$ < 40 $
&
$ < 44 $
&
$ < 36 $
&
$ < 40 $
&
$ < 44 $
\end{tabular}
\caption{ \label{tab:chr}
Limits at $1\sigma$,
$2\sigma$, and
$3\sigma$
for
the neutrino charge radii in units of $10^{-32} \, \text{cm}^2$,
obtained from the analysis of the COHERENT
CsI and Ar data,
and from the combined fit.
}
\end{center}
\end{table*}

\begin{figure*}[!t]
\centering
\setlength{\tabcolsep}{0pt}
\begin{tabular}{cc}
\subfigure[]{\label{fig:chr5-em-mt}
\begin{tabular}{c}
\includegraphics*[width=0.35\linewidth]{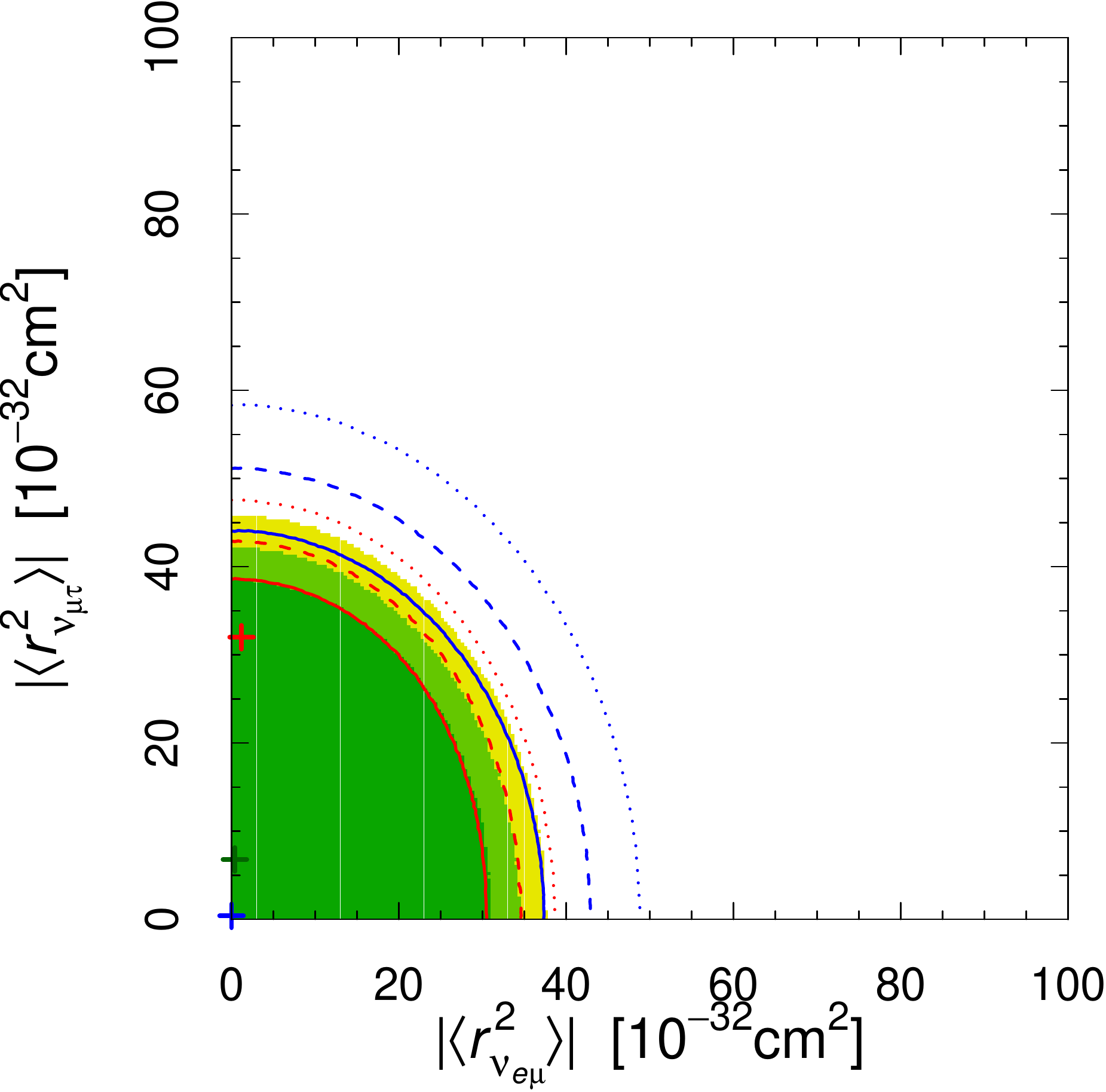}
\\
\end{tabular}
}
&
\subfigure[]{\label{fig:chr5-et-mt}
\begin{tabular}{c}
\includegraphics*[width=0.35\linewidth]{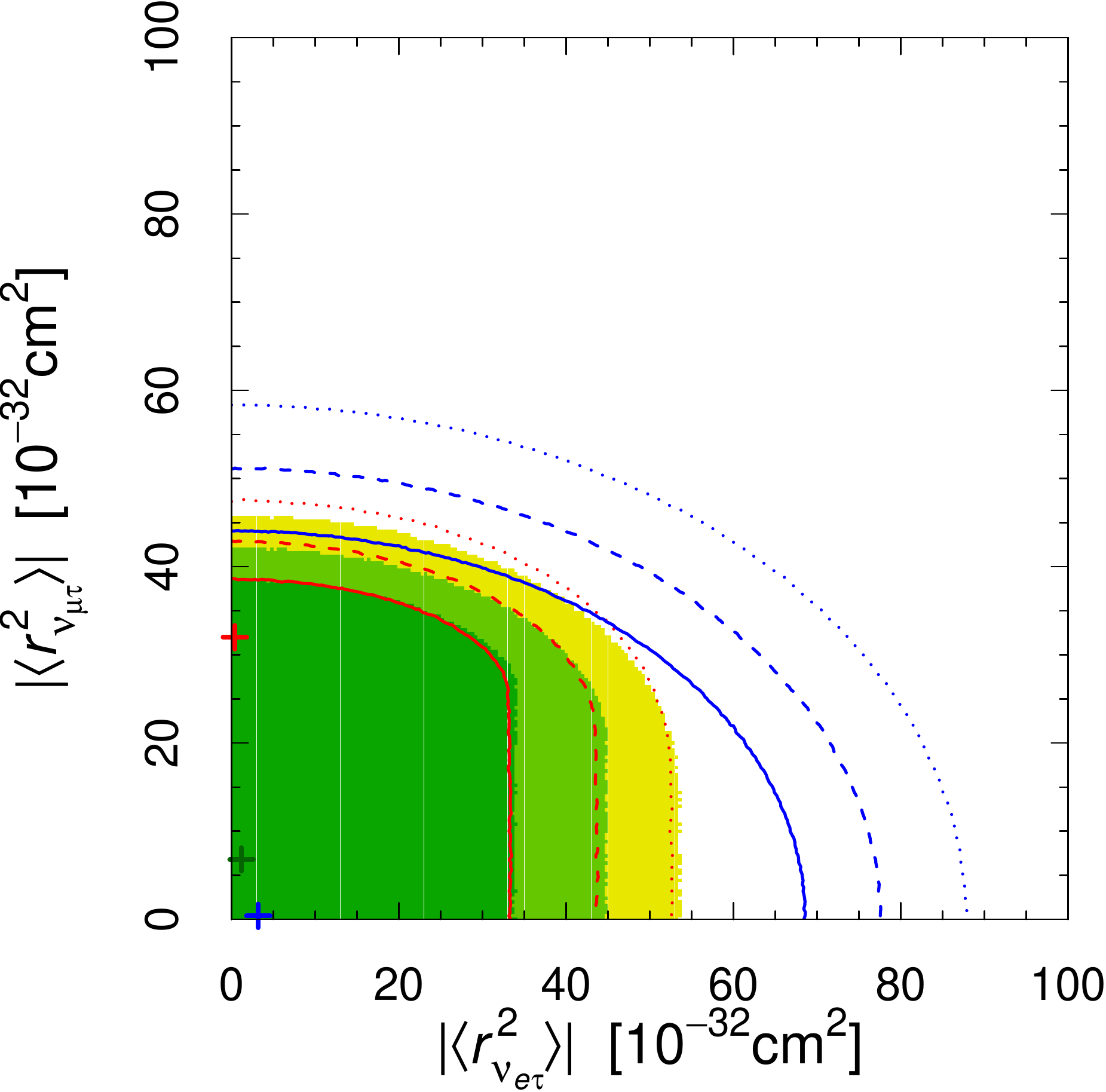}
\\
\end{tabular}
}
\\
\subfigure[]{\label{fig:chr5-em-et}
\begin{tabular}{c}
\includegraphics*[width=0.35\linewidth]{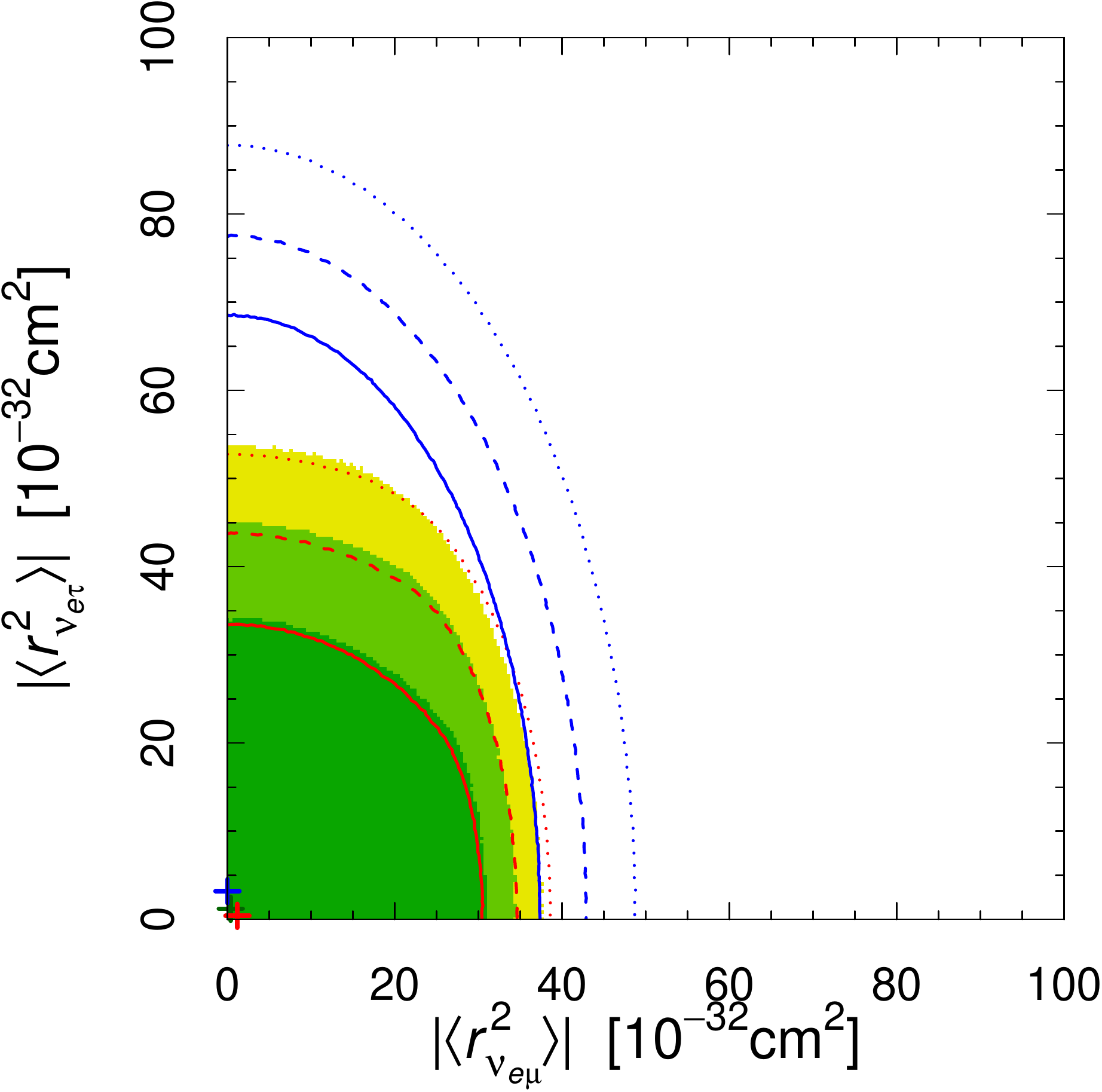}
\\
\end{tabular}
}
&
\subfigure[]{\label{fig:chr5-ee-mm}
\begin{tabular}{c}
\includegraphics*[width=0.35\linewidth]{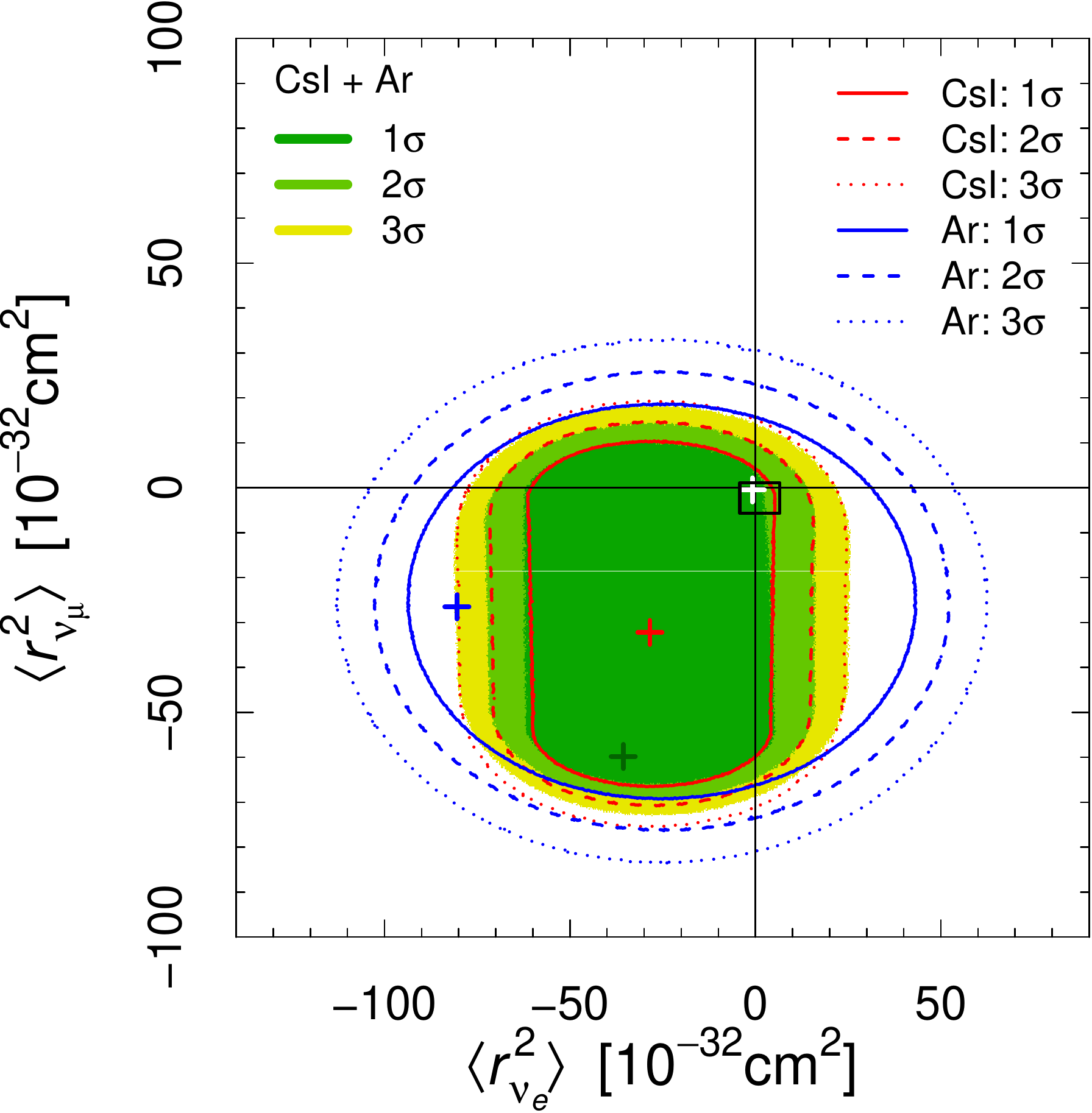}
\\
\end{tabular}
}
\end{tabular}
\caption{ \label{fig:chr5}
Contours of the allowed regions in different planes of the
neutrino charge radii parameter space
obtained with fixed $R_{n}$
obtained from the analysis of COHERENT
CsI data (red lines),
from the analysis of COHERENT
Ar data in this paper (blue lines),
and from the combined fit
(shaded green-yellow regions).
The crosses with the corresponding colors indicate the best fit points.
The white cross near the origin in panel \subref{fig:chr5-ee-mm} indicates the Standard Model values
in Eqs.~(\ref{reSM}) and (\ref{rmSM}).
The black rectangle near the origin shows the 90\% bounds on
$\langle{r}_{\nu_{e}}^2\rangle$
and
$\langle{r}_{\nu_{\mu}}^2\rangle$
obtained, respectively in the
TEXONO~\cite{Deniz:2009mu}
and
BNL-E734~\cite{Ahrens:1990fp}
experiments.
}
\end{figure*}

\begin{figure}[!t]
\centering
\includegraphics*[width=0.5\linewidth]{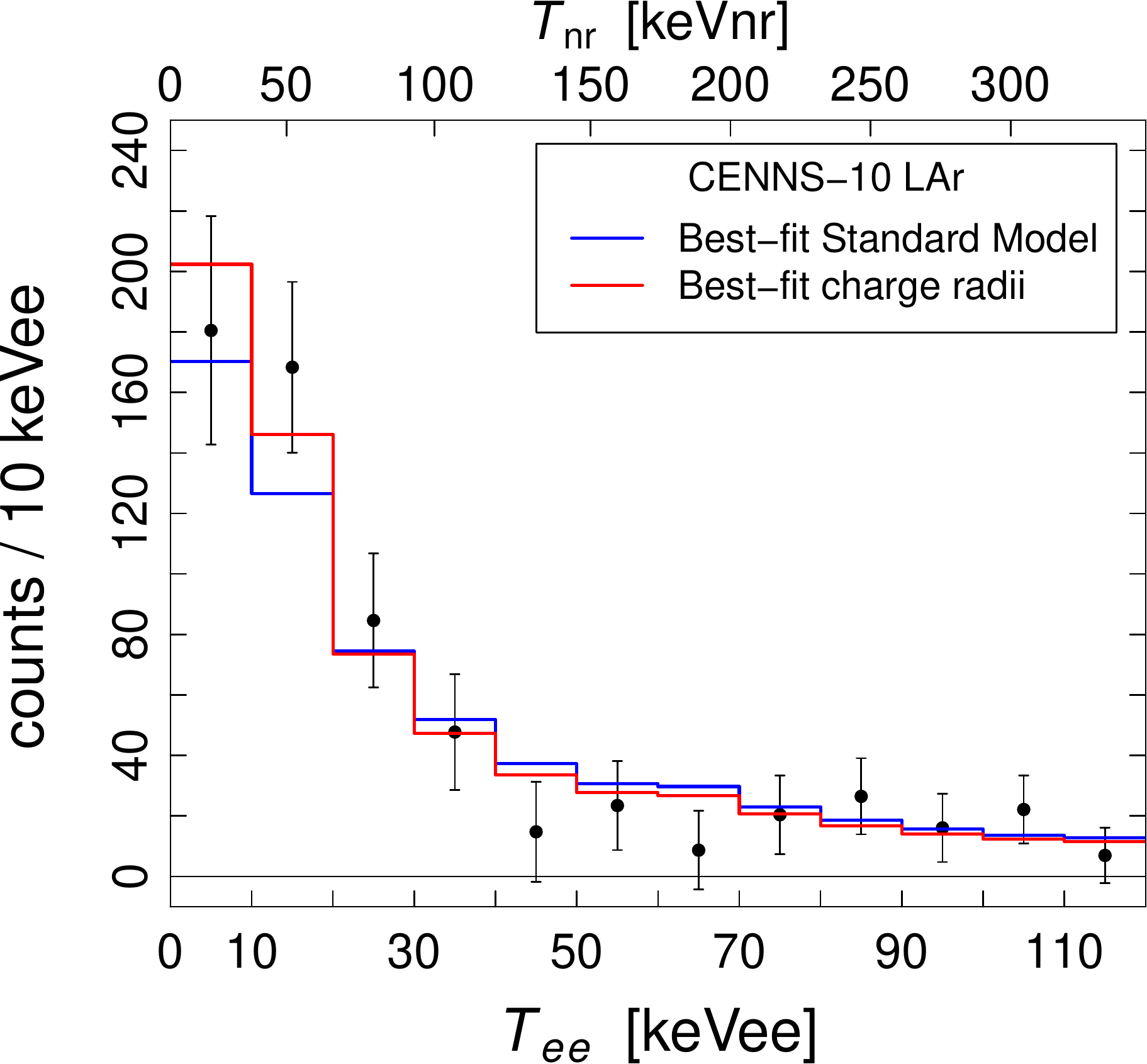}
\caption{ \label{fig:hist-chr}
Histogram representing the fits of the CENNS-10 data
(black points with statistical error bars)
with the Standard Model charge radii given in
Eqs.~(\ref{reSM}) and (\ref{rmSM})
(blue histogram),
and with the best-fit charge radii of the COHERENT Ar data analysis
(red histogram).
}
\end{figure}

\begin{figure}[!t]
\centering
\includegraphics*[width=0.5\linewidth]{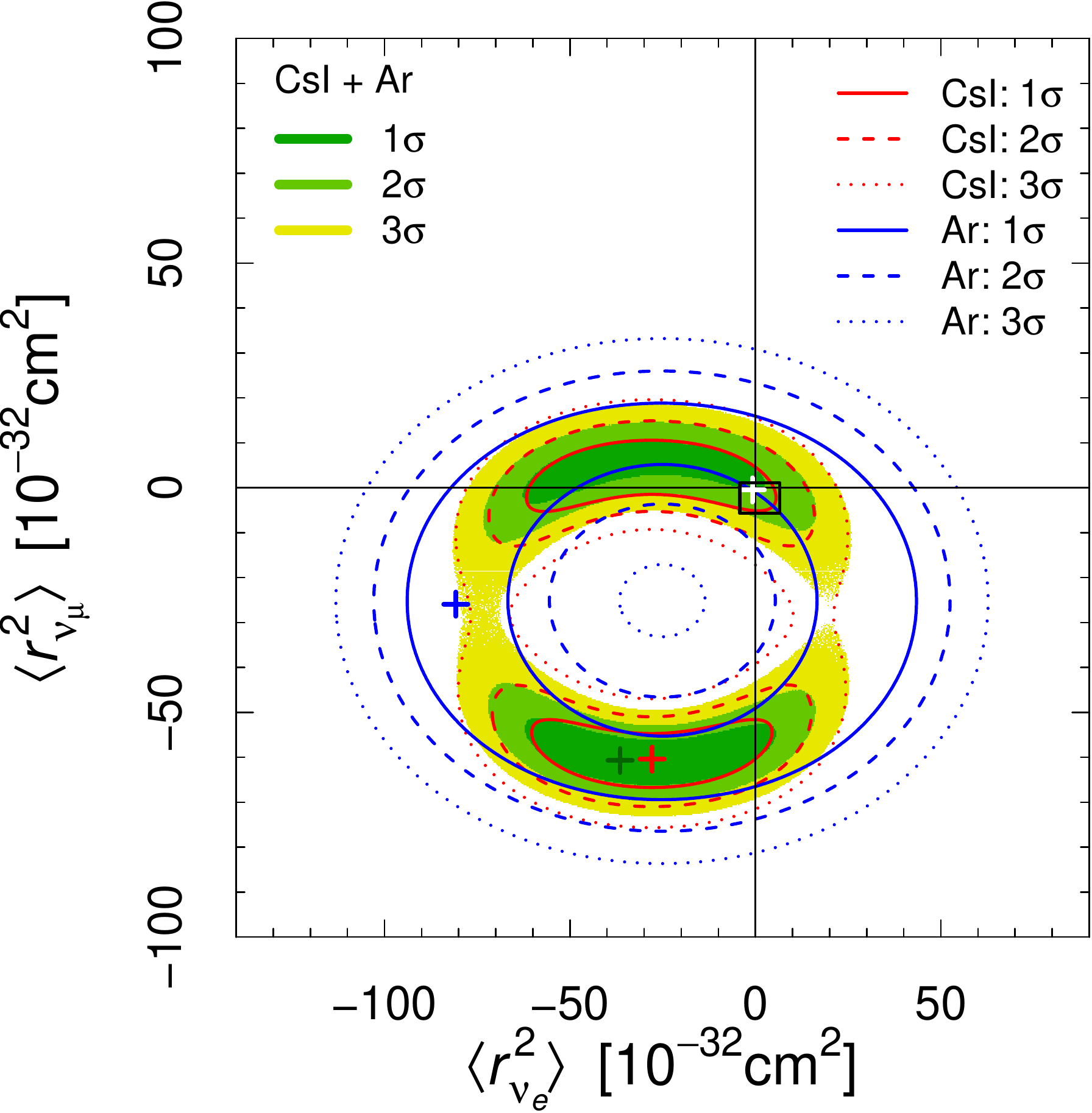}
\caption{ \label{fig:chr2}
Contours of the allowed regions in the
($\langle{r}_{\nu_{e}}^2\rangle,\langle{r}_{\nu_{\mu}}^2\rangle$)
plane
obtained with fixed $R_{n}$
obtained from the analysis of COHERENT
CsI data (red lines),
from the analysis of COHERENT
Ar data in this paper (blue lines),
and from the combined fit
(shaded green-yellow regions),
assuming the absence of transition charge radii.
The crosses with the corresponding colors indicate the best fit points.
The white cross near the origin indicates the Standard Model values
in Eqs.~(\ref{reSM}) and (\ref{rmSM}).
The black rectangle near the origin shows the 90\% bounds on
$\langle{r}_{\nu_{e}}^2\rangle$
and
$\langle{r}_{\nu_{\mu}}^2\rangle$
obtained, respectively in the
TEXONO~\cite{Deniz:2009mu}
and
BNL-E734~\cite{Ahrens:1990fp}
experiments.
}
\end{figure}

The results of our fits for fixed and free $R_{n}$ are given in Table~\ref{tab:chr}.
One can see that the bounds obtained with fixed and free $R_{n}$ are similar.
Therefore,
our results are practically independent from the unknown value of $R_{n}$,
and in the following, for simplicity,
we discuss only the case of fixed $R_{n}$.

The bounds in Table~\ref{tab:chr} obtained from the COHERENT Ar data
are compatible, but less stringent than those obtained from the CsI data,
and the bounds of the combined fit are similar to those obtained with the CsI data only.
This is illustrated by Figure~\ref{fig:chr5},
that depicts the allowed regions in different planes of the parameter space of the
neutrino charge radii.
It is interesting,
however,
that the contribution of the argon data shrinks the allowed region
in the vicinity of the Standard Model values
of the diagonal charge radii given in
Eqs.~(\ref{reSM}) and (\ref{rmSM})
and shown by the white cross near the origin in Figure~\ref{fig:chr5-ee-mm}.
In the combined fit,
the point corresponding to the Standard Model values
of the diagonal charge radii
lies at the edge of the $1\sigma$ allowed region.
The best fit of the COHERENT Ar data is obtained for
relatively large values of the charge radii
shown by the blue crosses in Figure~\ref{fig:chr5}.
As shown in Figure~\ref{fig:hist-chr},
the resulting enhancement of the CE$\nu$NS cross section
with respect to the SM
allows a better fit of the low-energy data,
while the medium- and high-energy data are fitted better with a slightly
lower background allowed by the uncertainties.
The best-fit large values of
$\langle{r}_{\nu_{e}}^2\rangle$
and
$\langle{r}_{\nu_{\mu}}^2\rangle$
are, however,
completely excluded by the bounds obtained by other experiments
(see Table~I of Ref.~\cite{Cadeddu:2018dux}).
The black rectangle near the origin in Figure~\ref{fig:chr5-ee-mm}
shows the most stringent 90\% bounds on
$\langle{r}_{\nu_{e}}^2\rangle$
and
$\langle{r}_{\nu_{\mu}}^2\rangle$
obtained, respectively in the
TEXONO~\cite{Deniz:2009mu}
and
BNL-E734~\cite{Ahrens:1990fp}
experiments.
Unfortunately the CE$\nu$NS data still do not allow us to
limit the neutrino charge radii with such small precision,
but it is interesting to see that they tend to favor
negative values of the charge radii.

We considered also the case of absence of the neutrino transition charge radii,
that is motivated by the attempt to probe the
values of the neutrino charge radii in the Standard Model,
where only the diagonal charge radii with the values in
Eqs.~(\ref{reSM}) and (\ref{rmSM}) exist.
It is also possible that the physics beyond the Standard Model generates
off-diagonal neutrino charge radii
that are much smaller than the diagonal charge radii
and can be neglected in a first approximation.
Figure~\ref{fig:chr2} shows the allowed regions
in the
($\langle{r}_{\nu_{e}}^2\rangle,\langle{r}_{\nu_{\mu}}^2\rangle$)
plane.
One can see that the contribution of the Ar data leads to a restriction
of the allowed regions.
Although the combined fit tends to favor the allowed island
at large negative values of $\langle{r}_{\nu_{\mu}}^2\rangle$,
we cannot consider it as possible,
because it lies well outside the
black rectangle near the origin that
shows the 90\% bounds of the
TEXONO~\cite{Deniz:2009mu}
and
BNL-E734~\cite{Ahrens:1990fp}
experiments.
The allowed island of the combined CsI and Ar analysis
for values of $\langle{r}_{\nu_{\mu}}^2\rangle$ around zero
is compatible at about $2\sigma$ with these bounds,
as well as with the Standard Model values of the neutrino charge radii.

\section{Neutrino electric charges}
\label{sec:charges}

As discussed in Ref.~\cite{Cadeddu:2018dux},
the CE$\nu$NS process is sensitive
not only to the neutrino charge radii,
but also to the neutrino electric charges.
Usually neutrinos are considered as exactly neutral particles,
but in theories beyond the SM they can have small electric charges
(often called millicharges).
This possibility was considered in many experimental and theoretical studies
(see the review in Ref.~\cite{Giunti:2014ixa}).

The differential CE$\nu$NS cross section
that takes into account the contribution of the neutrino electric charges
in addition to Standard Model neutral-current weak interactions is
\begin{equation}
\dfrac{d\sigma_{\nu_{\ell}\text{-}\mathcal{N}}}{d T_\mathrm{nr}}
(E,T_\mathrm{nr})
=
\dfrac{G_{\text{F}}^2 M}{\pi}
\left(
1 - \dfrac{M T_\mathrm{nr}}{2 E^2}
\right)
\left\{
\left[
\left( g_{V}^{p} - Q_{\ell\ell} \right)
Z
F_{Z}(|\vet{q}|^2)
+
g_{V}^{n}
N
F_{N}(|\vet{q}|^2)
\right]^2
+
Z^2
F_{Z}^2(|\vet{q}|^2)
\sum_{\ell'\neq\ell}
|Q_{\ell'\ell}|^2
\right\}
,
\label{cs-ech}
\end{equation}
with
$g_{V}^{p}$ and $g_{V}^{n}$
given, respectively,
by Eqs.~(\ref{gVpth}) and (\ref{gVnth}),
with the numerical values in Eqs.~(\ref{gVp-nue})--(\ref{gVn}).
The neutrino electric charges $q_{\nu_{\ell\ell'}}$ contribute
through~\cite{Kouzakov:2017hbc,Giunti:2014ixa}
\begin{equation}
Q_{\ell\ell'}
=
\dfrac{ 2 \sqrt{2} \pi \alpha }{ G_{\text{F}} q^2 }
\, q_{\nu_{\ell\ell'}}
,
\label{Qech}
\end{equation}
where $ q^2 = - 2 M T_{\mathrm{nr}} $ is the squared four-momentum transfer.
Although the electric charges of neutrinos and antineutrinos are opposite,
neutrinos and antineutrinos contribute with the same sign to the shift of
$\sin^2\!\vartheta_{W}$,
as in the case of the charge radii,
because also the weak neutral current couplings change sign
from neutrinos to antineutrinos.

In this Section, we present the bounds on the neutrino electric charges
that we obtained from the analysis of the
spectral Ar data of the COHERENT experiment~\cite{Akimov:2020pdx}
and those obtained with a combined fit of the CsI and Ar data.
We also revise the CsI limits on the electric charges presented
in Ref.~\cite{Cadeddu:2019eta}
because they have been obtained through an expression similar to
that in Eq.~(\ref{hatQchr})
(see Eq.~(30) of Ref.~\cite{Cadeddu:2019eta}),
that overestimates their contribution by about 10\%,
as discussed in Section~\ref{sec:radii} for the charge radii.

There are five electric charges that can be determined with the COHERENT
CE$\nu$NS data:
the two diagonal electric charges
$q_{\nu_{ee}}$
and
$q_{\nu_{\mu\mu}}$,
and the absolute values of the three transition electric charges
$q_{\nu_{e\mu}}=q_{\nu_{\mu e}}^{*}$,
$q_{\nu_{e\tau}}$, and
$q_{\nu_{\mu\tau}}$.

\begin{table*}[t!]
\begin{center}
\begin{tabular}{ccccccc}
\\
&
\multicolumn{3}{c}{Fixed $R_{n}$}
&
\multicolumn{3}{c}{Free $R_{n}$}
\\
&
$1\sigma$
&
$2\sigma$
&
$3\sigma$
&
$1\sigma$
&
$2\sigma$
&
$3\sigma$
\\
\hline
&
\multicolumn{6}{c}{CsI}
\\
$q_{\nu_{ee}}$
&
$ 0 \div 37 $
&
$ -13 \div 57 $
&
$ -24 \div 71 $
&
$ 0 \div 39 $
&
$ -15 \div 57 $
&
$ -27 \div 71 $
\\
$q_{\nu_{\mu\mu}}$
&
$ -8 \div 8 $
&
$ -13 \div 27 $
&
$ -19 \div 47 $
&
$ -8 \div 9 $
&
$ -14 \div 28 $
&
$ -20 \div 47 $
\\
$|q_{\nu_{e\mu}}|$
&
$ < 17 $
&
$ < 28 $
&
$ < 35 $
&
$ < 18 $
&
$ < 28 $
&
$ < 35 $
\\
$|q_{\nu_{e\tau}}|$
&
$ < 23 $
&
$ < 38 $
&
$ < 51 $
&
$ < 23 $
&
$ < 38 $
&
$ < 51 $
\\
$|q_{\nu_{\mu\tau}}|$
&
$ < 23 $
&
$ < 34 $
&
$ < 41 $
&
$ < 24 $
&
$ < 34 $
&
$ < 41 $
\\
&
\multicolumn{6}{c}{Ar}
\\
$q_{\nu_{ee}}$
&
$ -17 \div 18 $
&
$ -23 \div 38 $
&
$ -28 \div 47 $
&
$ -16 \div 18 $
&
$ -23 \div 38 $
&
$ -28 \div 47 $
\\
$q_{\nu_{\mu\mu}}$
&
$ -8 \div 14 $
&
$ -11 \div 28 $
&
$ -15 \div 35 $
&
$ -7 \div 14 $
&
$ -11 \div 28 $
&
$ -15 \div 35 $
\\
$|q_{\nu_{e\mu}}|$
&
$ < 12 $
&
$ < 18 $
&
$ < 21 $
&
$ < 12 $
&
$ < 17 $
&
$ < 21 $
\\
$|q_{\nu_{e\tau}}|$
&
$ < 22 $
&
$ < 32 $
&
$ < 38 $
&
$ < 21 $
&
$ < 32 $
&
$ < 38 $
\\
$|q_{\nu_{\mu\tau}}|$
&
$ < 14 $
&
$ < 21 $
&
$ < 25 $
&
$ < 14 $
&
$ < 21 $
&
$ < 25 $
\\
&
\multicolumn{6}{c}{CsI + Ar}
\\
$q_{\nu_{ee}}$
&
$ -4 \div 24 $
&
$ -14 \div 34 $
&
$ -20 \div 42 $
&
$ -5 \div 23 $
&
$ -14 \div 34 $
&
$ -20 \div 41 $
\\
$q_{\nu_{\mu\mu}}$
&
$ -7 \div 4 $
&
$ -10 \div 12 $
&
$ -12 \div 20 $
&
$ -7 \div 3 $
&
$ -10 \div 12 $
&
$ -13 \div 20 $
\\
$|q_{\nu_{e\mu}}|$
&
$ < 11 $
&
$ < 17 $
&
$ < 20 $
&
$ < 11 $
&
$ < 16 $
&
$ < 20 $
\\
$|q_{\nu_{e\tau}}|$
&
$ < 18 $
&
$ < 27 $
&
$ < 34 $
&
$ < 17 $
&
$ < 27 $
&
$ < 33 $
\\
$|q_{\nu_{\mu\tau}}|$
&
$ < 14 $
&
$ < 20 $
&
$ < 25 $
&
$ < 14 $
&
$ < 20 $
&
$ < 24 $
\end{tabular}
\caption{ \label{tab:ech}
Limits at $1\sigma$,
$2\sigma$, and
$3\sigma$
for
the neutrino electric charges in units of $10^{-8} \, e$,
obtained from the analysis of the COHERENT
CsI and Ar data,
and from the combined fit.
}
\end{center}
\end{table*}

\begin{figure*}[!t]
\centering
\setlength{\tabcolsep}{0pt}
\begin{tabular}{cc}
\subfigure[]{\label{fig:ech5-em-mt}
\begin{tabular}{c}
\includegraphics*[width=0.35\linewidth]{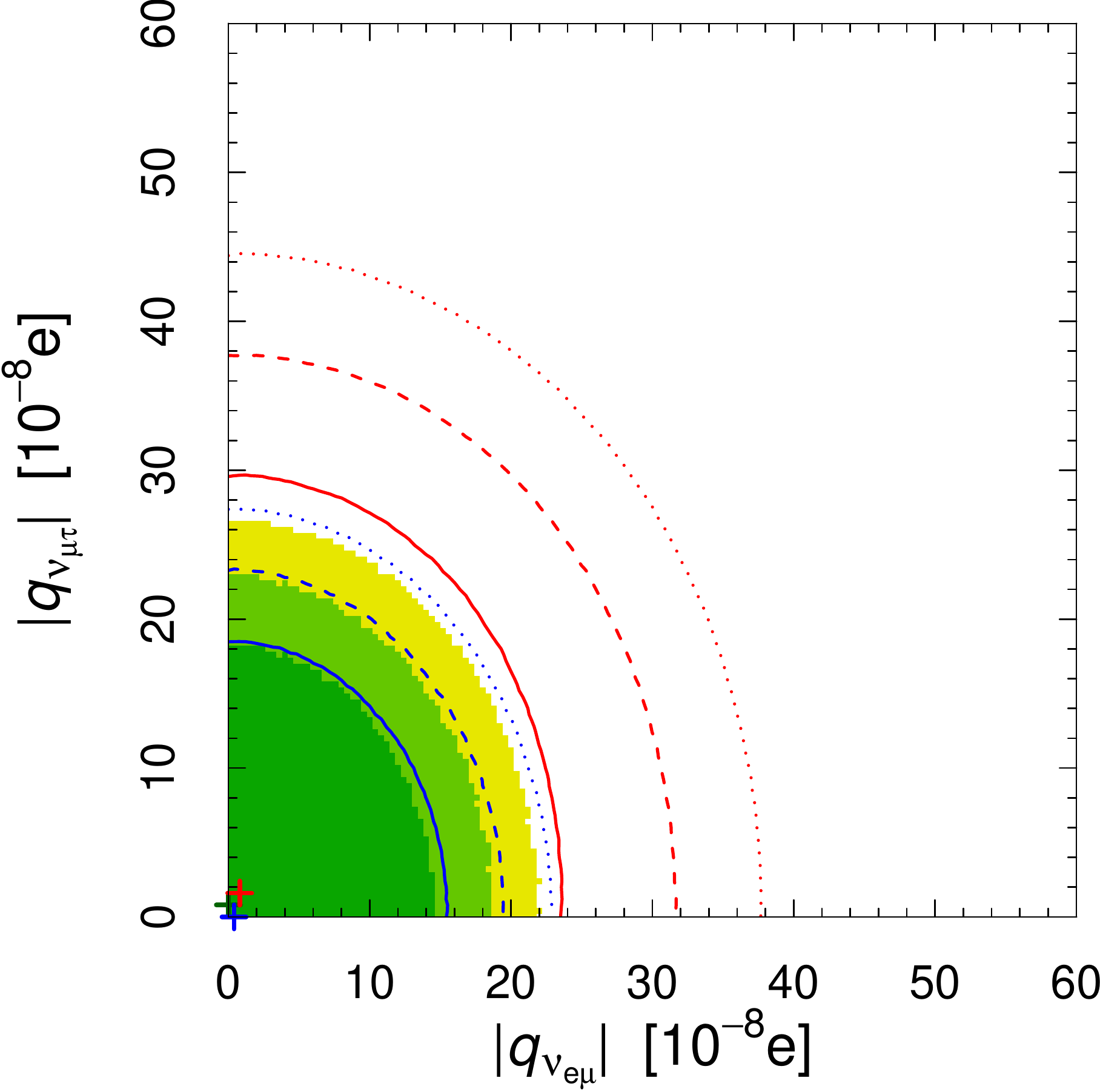}
\\
\end{tabular}
}
&
\subfigure[]{\label{fig:ech5-et-mt}
\begin{tabular}{c}
\includegraphics*[width=0.35\linewidth]{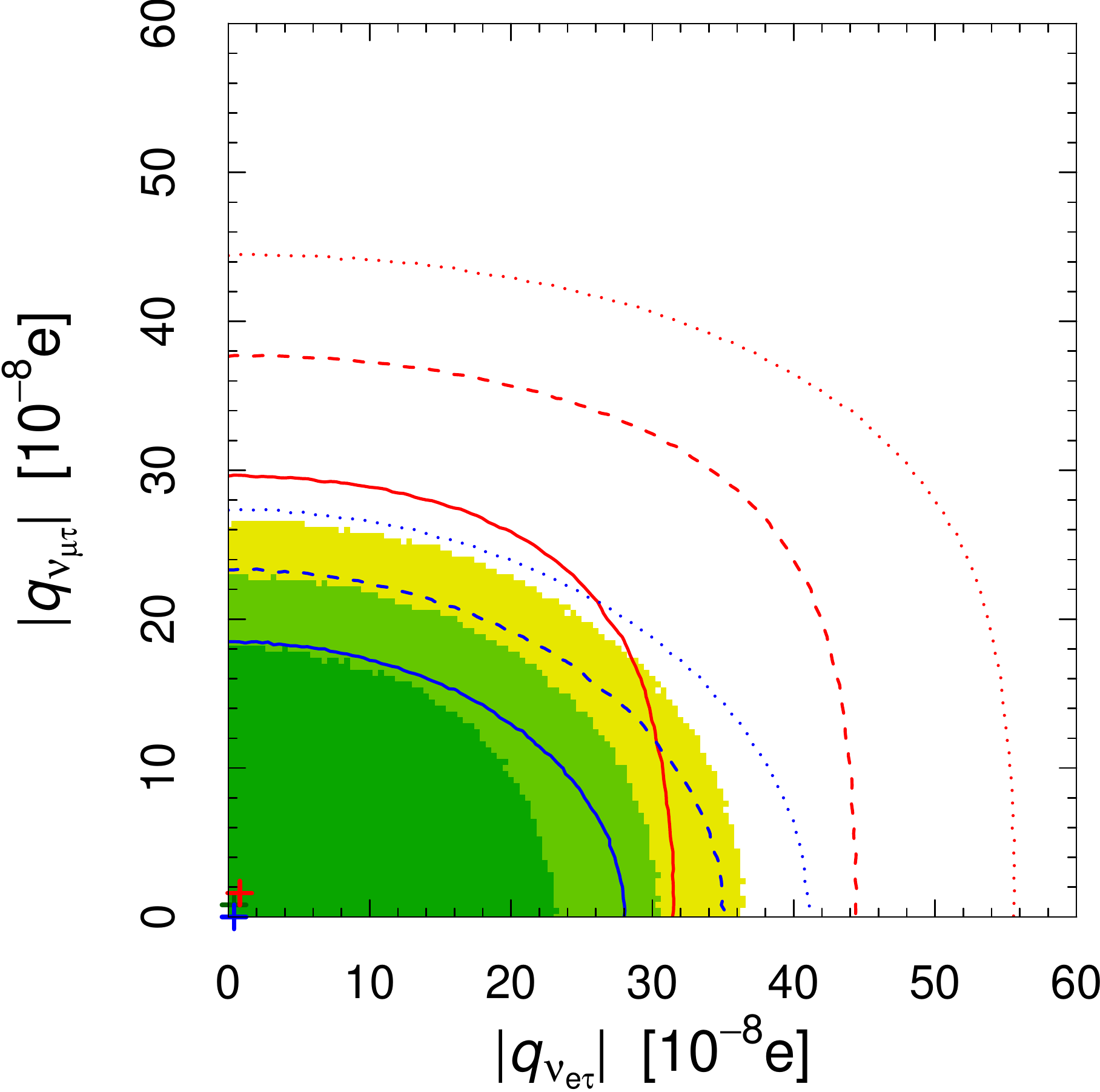}
\\
\end{tabular}
}
\\
\subfigure[]{\label{fig:ech5-em-et}
\begin{tabular}{c}
\includegraphics*[width=0.35\linewidth]{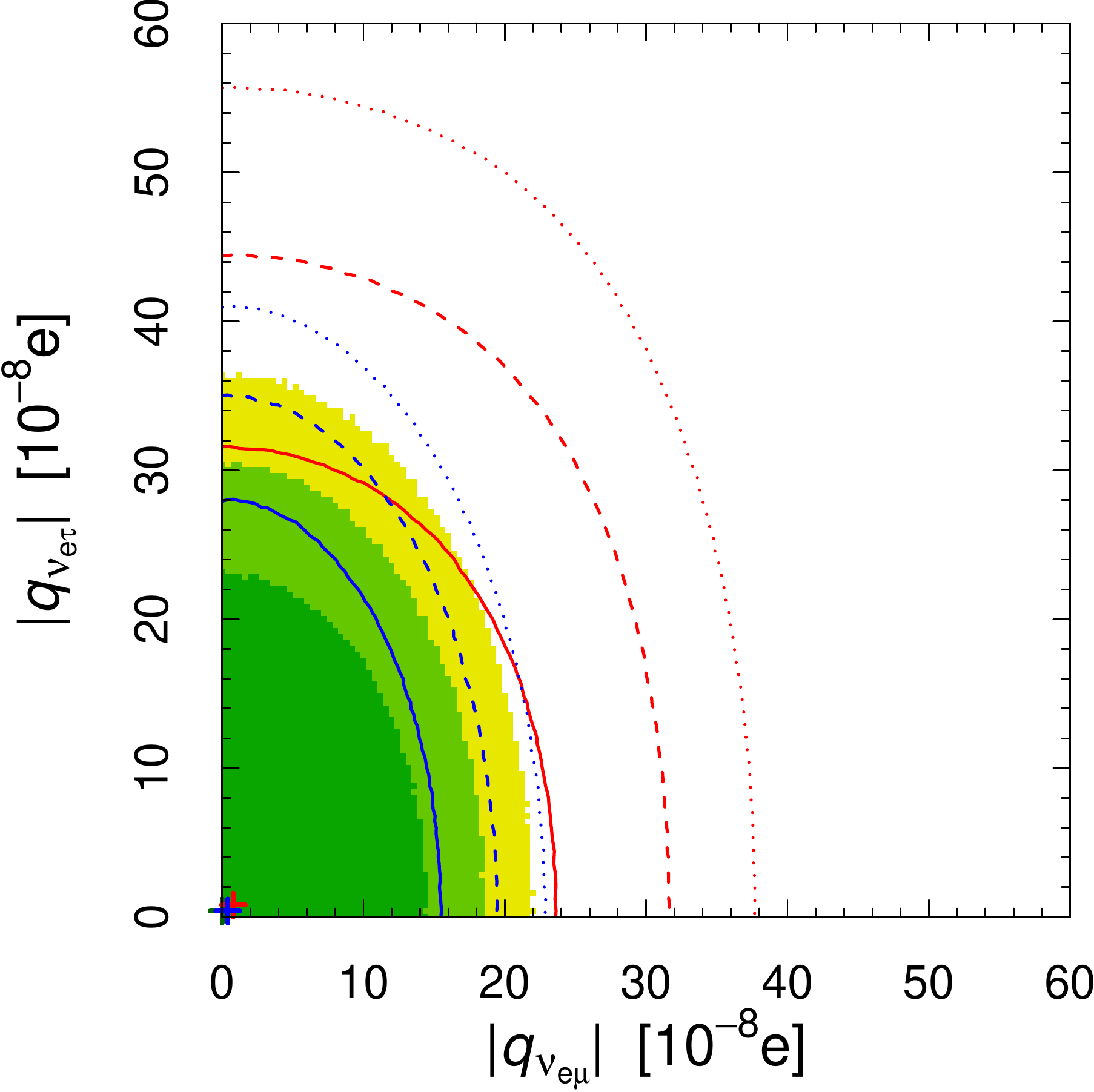}
\\
\end{tabular}
}
&
\subfigure[]{\label{fig:ech5-ee-mm}
\begin{tabular}{c}
\includegraphics*[width=0.35\linewidth]{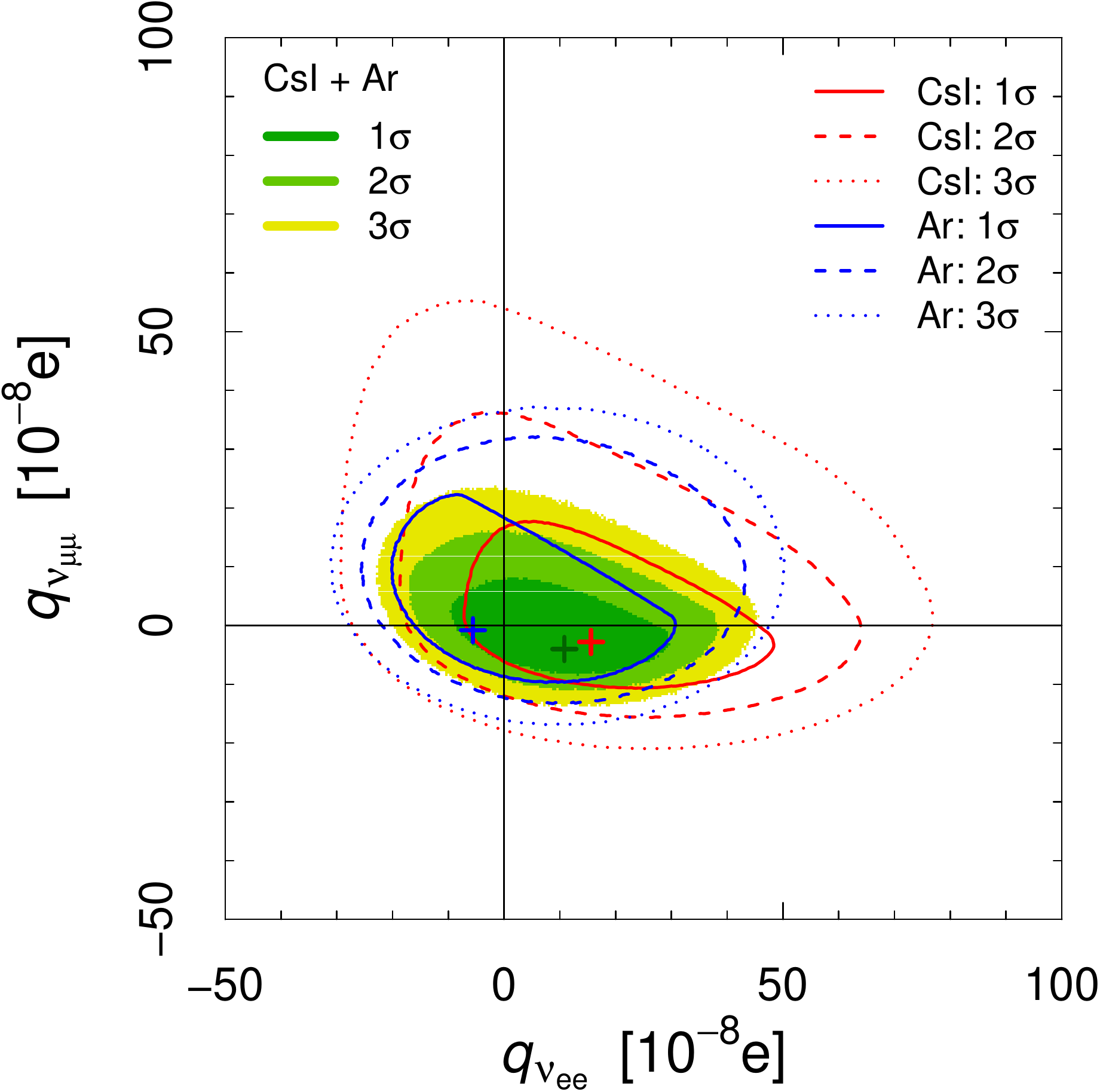}
\\
\end{tabular}
}
\end{tabular}
\caption{ \label{fig:ech5}
Contours of the allowed regions in different planes of the
neutrino electric charge parameter space
obtained with fixed $R_{n}$
obtained from the analysis of COHERENT
CsI data (red lines),
from the analysis of COHERENT
Ar data in this paper (blue lines),
and from the combined fit
(shaded green-yellow regions).
The crosses with the corresponding colors indicate the best fit points.
}
\end{figure*}

The results of our fits for fixed and free $R_{n}$ are given in Table~\ref{tab:ech}.
Since the bounds are similar in the two cases,
in Figure~\ref{fig:ech5}
we show only the allowed regions in different planes of the
neutrino electric charge parameter space obtained with fixed $R_{n}$.

From Table~\ref{tab:ech} and Figure~\ref{fig:ech5}
one can see that the COHERENT Ar data allow us to put
slightly more stringent limits
on the neutrino electric charges than the CsI data,
in spite of the larger uncertainties.
The larger sensitivity of the Ar data to the electric charges
is in contrast with the smaller sensitivity
to the charge radii discussed in Section~\ref{sec:radii}.
It follows from the enhancement of the neutrino electric charge effect
in CE$\nu$NS at low $q^2$,
because of the denominator in Eq.~(\ref{Qech}).
Since $ q^2 = - 2 M T_{\mathrm{nr}} $,
light nuclei are more sensitive than heavier ones
at the neutrino electric charges
for similar nuclear recoil kinetic energies $T_{\mathrm{nr}}$.
The acceptance functions of both the CsI and Ar experiments
have a threshold of about 5 $\mathrm{keV}_{\mathrm{nr}}$.
Since
$ M(^{40}\text{Ar}) \simeq 37 \, \text{GeV}$,
$ M(^{133}\text{Cs}) \simeq 123 \, \text{GeV}$, and
$ M(^{127}\text{I}) \simeq 118 \, \text{GeV}$,
the minimum value of $|q^2|$
can be about 3.2 times smaller in the Ar experiment
than in the CsI experiment.
However,
this enhancement of a factor as large as 3.2
of the neutrino electric charge effect for nuclear recoil kinetic energies
above the experimental threshold is mitigated by the different
sizes of the energy bins:
in the Ar experiment the first bin includes energies from the threshold
to about 36 $\mathrm{keV}_{\mathrm{nr}}$,
whereas the CsI energy bins have a size of about 1.7 $\mathrm{keV}_{\mathrm{nr}}$.
Therefore, the enhancement of the electric charge effect
occurs only in the first energy bin of the Ar experiment.
Nevertheless, this enhancement is sufficient for achieving
a slightly better performance of the Ar data
in constraining the neutrino electric charges
in spite of the larger uncertainties,
as can be seen in Table~\ref{tab:ech} and Figure~\ref{fig:ech5}.

The combined fit of the COHERENT CsI and Ar data
leads to a significant restriction of the allowed values of
the neutrino electric charges,
especially the diagonal ones,
because of the incomplete overlap of the
CsI and Ar allowed regions that can be seen in
Figure~\ref{fig:ech5-ee-mm}.
Although the best-fit values of
$q_{\nu_{ee}}$
and
$q_{\nu_{\mu\mu}}$
are visibly different from zero,
the deviation is not significant, because
the $1\sigma$ allowed region includes well the point
$q_{\nu_{ee}}=q_{\nu_{\mu\mu}}=0$.
From Figures~\ref{fig:ech5-em-mt},
\ref{fig:ech5-et-mt}, and
\ref{fig:ech5-em-et},
one can see that the best-fit values of the off-diagonal electric charges
are close to zero and the values of the off-diagonal electric charges
are well constrained.

As already noted in Ref.~\cite{Cadeddu:2019eta},
the bounds of the order of $10^{-7} \, e$
that we obtained are not competitive with the bounds on the
electron neutrino electric charges obtained in reactor neutrino experiments,
that are at the level of
$10^{-12} \, e$~\cite{Studenikin:2013my,Giunti:2014ixa,Chen:2014dsa,Tanabashi:2018oca}.
These limits are given in the literature for the diagonal
electron neutrino charge
$q_{\nu_{ee}}$,
because the contribution of the off-diagonal charges was not considered.
However, since the off-diagonal charges contribute to the
cross section in a quantitatively comparable way, we can consider
them to be bounded at the same order of magnitude level of
$10^{-12} \, e$.
Therefore our bounds are not competitive with the reactor bounds for
$q_{\nu_{ee}}$,
$q_{\nu_{e\mu}}$, and
$q_{\nu_{e\tau}}$.
On the other hand,
they are the only existing laboratory bounds for
$q_{\nu_{\mu\mu}}$ and $q_{\nu_{\mu\tau}}$.

\section{Neutrino magnetic moments}
\label{sec:magnetic}

The neutrino magnetic moment is the electromagnetic
neutrino property that is most studied and searched experimentally.
The reason is that its existence is predicted by many
models beyond the Standard Model,
especially those that include right-handed neutrinos.
It is also phenomenologically important for astrophysics
because neutrinos with a magnetic moment can interact
with astrophysical magnetic fields leading to several
important effects
(see the reviews in Refs.~\cite{Giunti:2014ixa,Giunti:2015gga}).

The CE$\nu$NS process is sensitive to neutrino magnetic
moments~\cite{Kosmas:2017tsq,Papoulias:2019txv,Khan:2019mju,Cadeddu:2019eta,Miranda:2019wdy,Papoulias:2019xaw}.
In this Section, we present the bounds on the neutrino magnetic moments
that we obtained from the analysis of the COHERENT Ar data
and those that we obtained from the combined fit of the
COHERENT CsI and Ar data.

For the analysis of the COHERENT data we used the least-squares function in Eq.~(\ref{chi-spectrum}),
with the theoretical predictions $N_i^{\mathrm{CE\nu NS}}$
calculated by adding to the Standard Model weak cross section in Eq.~(\ref{cs-std})
the magnetic moment interaction cross section
\begin{equation}
\dfrac{d\sigma_{\nu_{\ell}\text{-}\mathcal{N}}^{\text{mag}}}{d T_\mathrm{nr}}
(E,T_\mathrm{nr})
=
\dfrac{ \pi \alpha^2 }{ m_{e}^2 }
\left( \dfrac{1}{T_\mathrm{nr}} - \dfrac{1}{E} \right)
Z^2 F_{Z}^2(|\vet{q}|^2)
\left| \dfrac{\mu_{\nu_{\ell}}}{\mu_{\text{B}}} \right|^2
,
\label{cs-mag}
\end{equation}
where $m_{e}$ is the electron neutrino mass,
$\mu_{\nu_{\ell}}$
is the effective magnetic moment of the flavor neutrino $\nu_{\ell}$
in elastic scattering
(see Ref.~\cite{Giunti:2014ixa}),
and $\mu_{\text{B}}$ is the Bohr magneton.

\begin{table*}[t!]
\begin{center}
\begin{tabular}{ccccccc}
\\
&
\multicolumn{3}{c}{Fixed $R_{n}$}
&
\multicolumn{3}{c}{Free $R_{n}$}
\\
&
$1\sigma$
&
$2\sigma$
&
$3\sigma$
&
$1\sigma$
&
$2\sigma$
&
$3\sigma$
\\
\hline
&
\multicolumn{6}{c}{CsI}
\\
$|\mu_{\nu_{e}}|$
&
$ < 24 $
&
$ < 42 $
&
$ < 58 $
&
$ < 33 $
&
$ < 50 $
&
$ < 65 $
\\
$|\mu_{\nu_{\mu}}|$
&
$ < 26 $
&
$ < 34 $
&
$ < 42 $
&
$ 3 \div 31 $
&
$ < 39 $
&
$ < 46 $
\\
&
\multicolumn{6}{c}{Ar}
\\
$|\mu_{\nu_{e}}|$
&
$ < 55 $
&
$ < 70 $
&
$ < 85 $
&
$ < 55 $
&
$ < 70 $
&
$ < 85 $
\\
$|\mu_{\nu_{\mu}}|$
&
$ < 39 $
&
$ < 50 $
&
$ < 60 $
&
$ < 39 $
&
$ < 50 $
&
$ < 60 $
\\
&
\multicolumn{6}{c}{CsI + Ar}
\\
$|\mu_{\nu_{e}}|$
&
$ < 27 $
&
$ < 44 $
&
$ < 56 $
&
$ < 33 $
&
$ < 48 $
&
$ < 60 $
\\
$|\mu_{\nu_{\mu}}|$
&
$ 5 \div 27 $
&
$ < 34 $
&
$ < 41 $
&
$ 12 \div 31 $
&
$ < 37 $
&
$ < 43 $
\end{tabular}
\caption{ \label{tab:mag}
Limits at $1\sigma$,
$2\sigma$, and
$3\sigma$ for
the neutrino magnetic moments in units of $10^{-10} \, \mu_{\text{B}}$,
obtained from the analysis of COHERENT
CsI data in Ref.~\cite{Cadeddu:2019eta},
from the analysis of COHERENT
Ar data in this paper,
and from the combined fit.
}
\end{center}
\end{table*}

\begin{figure*}[!t]
\centering
\setlength{\tabcolsep}{0pt}
\begin{tabular}{cc}
\subfigure[]{\label{fig:mag2}
\begin{tabular}{c}
\includegraphics*[width=0.49\linewidth]{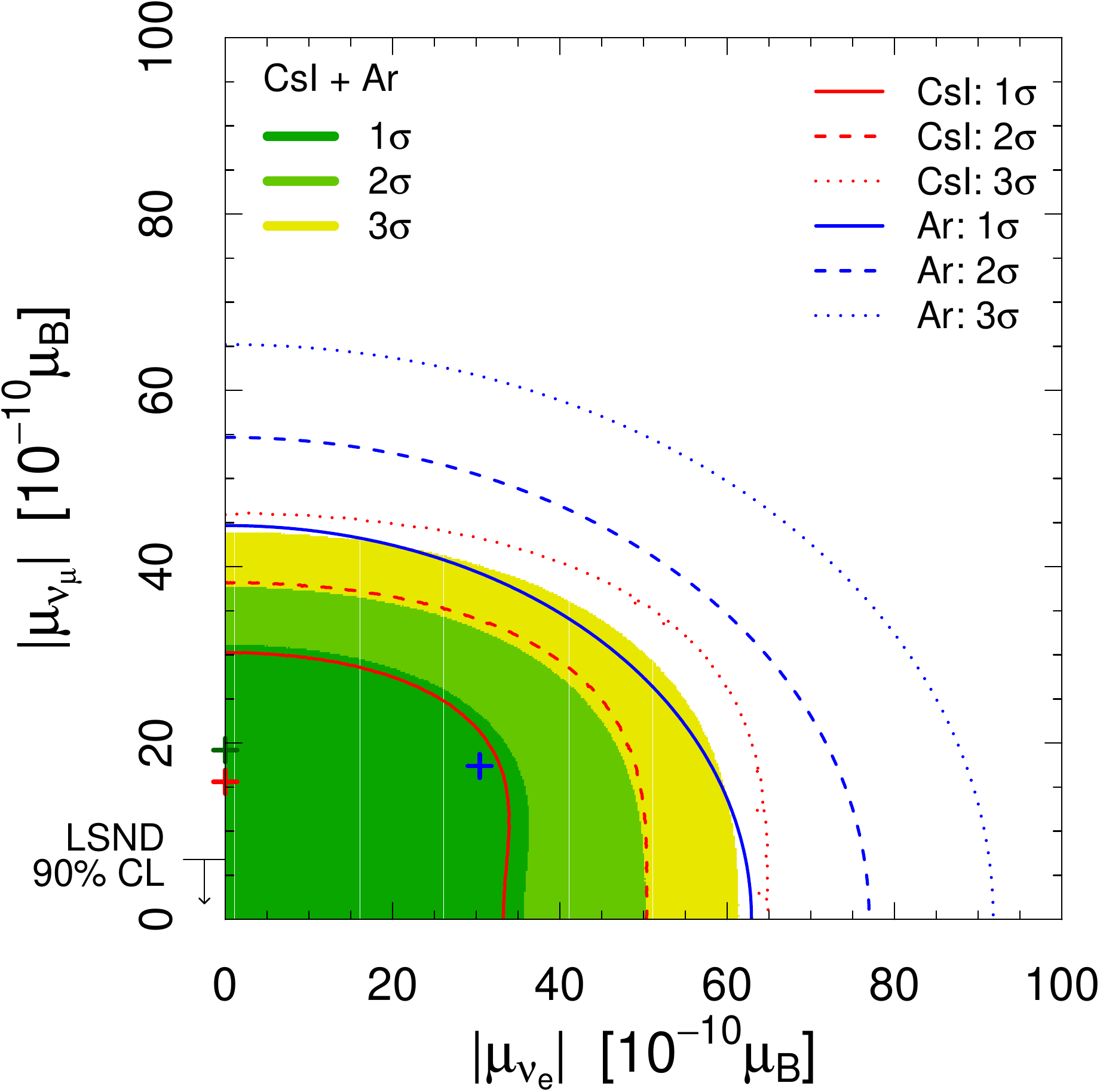}
\\
\end{tabular}
}
&
\subfigure[]{\label{fig:hist-mag}
\begin{tabular}{c}
\includegraphics*[width=0.49\linewidth]{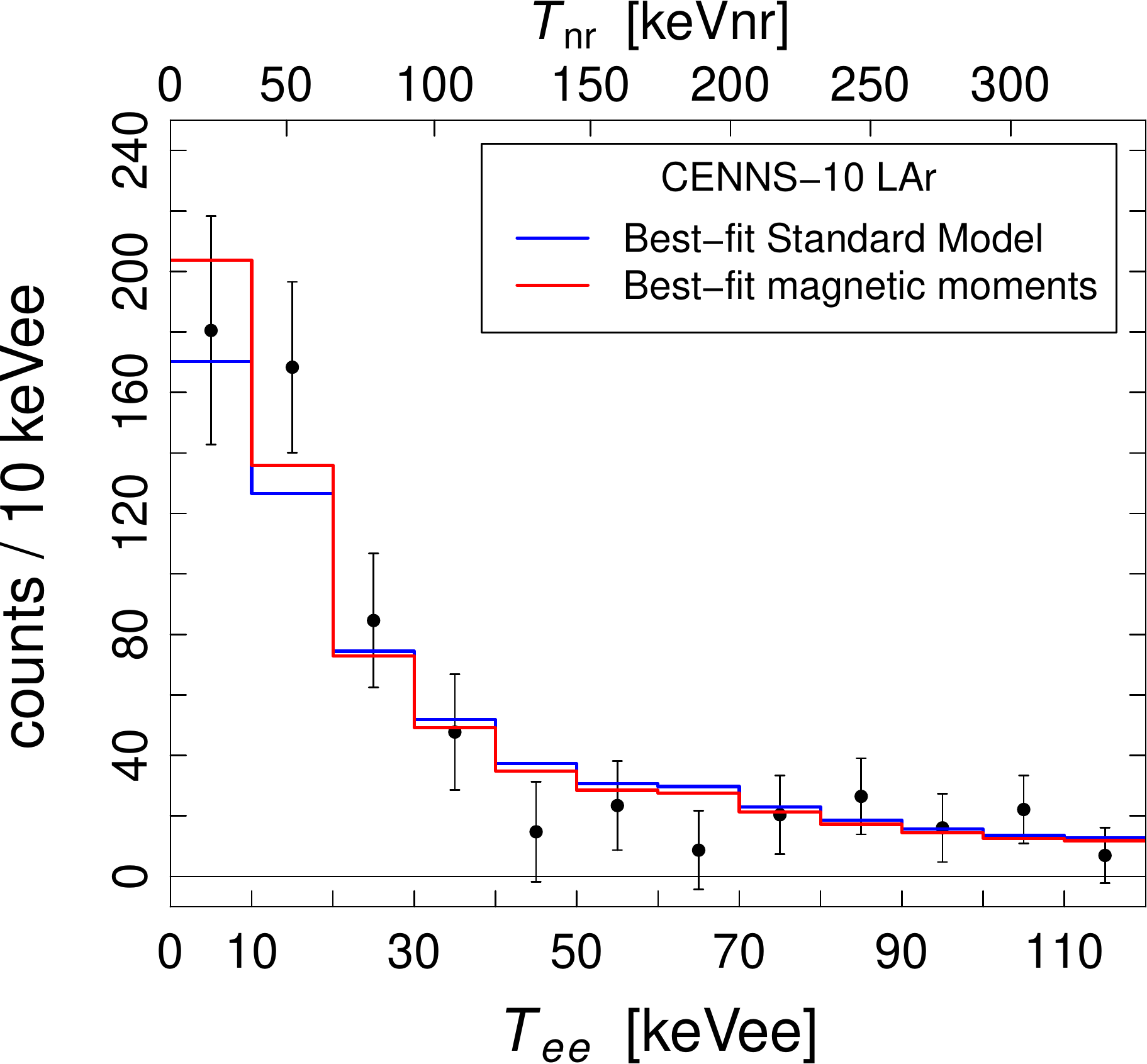}
\\
\end{tabular}
}
\end{tabular}
\caption{ \label{fig:mag}
\subref{fig:mag2}
Contours of the allowed regions in the
($|\mu_{\nu_{e}}|,|\mu_{\nu_{\mu}}|$) plane
obtained with fixed $R_{n}$
obtained from the analysis of COHERENT
CsI data in Ref.~\cite{Cadeddu:2019eta} (red lines),
from the analysis of COHERENT
Ar data in this paper (blue lines),
and from the combined fit
(shaded green-yellow regions).
The crosses with the corresponding colors indicate the best fit points.
The figure shows also the LSND 90\% CL upper bound
on $|\mu_{\nu_{\mu}}|$~\cite{Auerbach:2001wg}.
\subref{fig:hist-mag}
Histograms representing the fits of the CENNS-10 data
(black points with statistical error bars)
with the Standard Model  weak-interaction cross section
(blue histogram),
and with the best-fit magnetic moment of the COHERENT Ar data analysis
(red histogram).
}
\end{figure*}

The results of the fits for fixed and free $R_{n}$ are given in Table~\ref{tab:mag}.
Again,
one can see that the bounds are robust with respect to our lack of knowledge of the
value of $R_{n}$,
because the bounds are similar for fixed and free $R_{n}$.
For simplicity, in Figure~\ref{fig:mag2}
we show only the allowed regions in the
($|\mu_{\nu_{e}}|,|\mu_{\nu_{\mu}}|$) plane obtained with fixed $R_{n}$.

From Figure~\ref{fig:mag2} one can see that the best fit of the Ar data
is obtained for relatively large values of the neutrino magnetic moments.
The reason is similar to that discussed in Section~\ref{sec:radii}
for the neutrino charge radii:
as illustrated in Figure~\ref{fig:hist-mag},
the enhancement of the CE$\nu$NS cross section with
a sizable neutrino magnetic moment contribution
fits better the low-energy bins of the Ar data set
than the SM cross section
and the medium- and high-energy bins are fitted better with
a slightly smaller background allowed by the uncertainties.
In the combined CsI and Ar analysis we find the best fit for
$|\mu_{\nu_{e}}|=0$,
but a best-fit value of $|\mu_{\nu_{\mu}}|$ that is relatively large.
However,
we cannot consider this as a valid indication in favor of
a non-zero $|\mu_{\nu_{\mu}}|$ because the best-fit value is much larger than the
bounds obtained in accelerator experiments with
$\nu_{\mu}-e$ scattering
(see Table~IV of Ref.~\cite{Giunti:2014ixa}).
The most stringent of those bounds is the LSND bound
$ |\mu_{\nu_{\mu}}| < 6.8 \times 10^{-10} \, \mu_{\text{B}}$
at 90\% CL~\cite{Auerbach:2001wg}
shown in Figure~\ref{fig:mag2}.
Nevertheless,
the $1\sigma$ allowed region of the combined fit is compatible with this bound,
as well as with the stringent bounds on $|\mu_{\nu_{e}}|$
established in reactor neutrino experiments
(the currently best one,
$
|\mu_{\nu_{e}}| < 2.9 \times 10^{-11} \, \mu_{\text{B}}
$~\cite{Beda:2012zz,Giunti:2014ixa,Tanabashi:2018oca},
is not shown in Figure~\ref{fig:mag2} because it would not be
distinguishable from the $y$ axis).

\section{Conclusions}
\label{sec:conclusions}

In this paper we discussed the information on
nuclear physics, on the low-energy electroweak mixing angle
and on the electromagnetic properties of neutrinos
that can be obtained from the analysis of the recent
CE$\nu$NS data on argon of the COHERENT experiment~\cite{Akimov:2020pdx}.
We also presented the results obtained by combining the analysis
of the COHERENT Ar data with the analysis of the COHERENT CsI data~\cite{Akimov:2017ade}
performed in Ref.~\cite{Cadeddu:2019eta}.

The information on nuclear physics provided by CE$\nu$NS measurements
concerns the radius of the neutron distribution in the target nucleus.
In Section~\ref{sec:neutron} we calculated the bounds
on the radius of the neutron distribution in $^{40}\text{Ar}$.
These bounds are in agreement with the nuclear model predictions
in Table~\ref{tab:models},
but are rather weak,
because the data have large uncertainties.
Therefore, they do not allow us to discriminate the different nuclear models.

For the low-energy weak mixing angle,
from the analysis of the COHERENT Ar data
we obtained a relatively large value which,
however, is compatible with that predicted by the Standard Model
at about $1.7\sigma$.
Including in the analysis the COHERENT CsI data, we found a
value that is still larger than that predicted by the Standard Model,
but compatible at about $1\sigma$.

The analysis of the COHERENT Ar data allows us to constrain
the neutrino charge radii and magnetic moments,
but not as well as the analysis of the COHERENT CsI data.
Therefore,
the combined fits are dominated by the CsI data,
with small changes due to the Ar data with respect to the results obtained in
Ref.~\cite{Cadeddu:2019eta}.
On the other hand,
the Ar data are more sensitive to the neutrino electric charges
than the CsI data because of the lower nuclear mass,
as discussed in Section~\ref{sec:charges}.
Therefore,
the Ar data allowed us to improve the constraints on the neutrino electric charges
that can be obtained with CE$\nu$NS.
In particular, we improved the only existing
laboratory bounds on the electric charge $q_{\mu\mu}$ of the muon neutrino
and on the transition electric charge $q_{\mu\tau}$.

In conclusion, we would like to emphasize the importance of the
results of the COHERENT experiment,
that opened the way for CE$\nu$NS measurements, first with the
CsI detector~\cite{Akimov:2017ade}
and then with the LAr detector~\cite{Akimov:2020pdx}.
Even if the first CE$\nu$NS data on argon have large uncertainties,
they give us useful physical information.
We believe that future experimental improvements will lead to
far-reaching results.

\begin{acknowledgments}
The work of C. Giunti was partially supported by the research grant "The Dark Universe: A Synergic Multimessenger Approach" number 2017X7X85K under the program PRIN 2017 funded by the Ministero dell'Istruzione, Universit\`a e della Ricerca (MIUR).
The work of Y.F. Li and Y.Y. Zhang is supported by the National Natural Science Foundation of China under Grant No. 11835013. Y.F. Li is also grateful for the support by the CAS Center for Excellence in Particle Physics (CCEPP).
\end{acknowledgments}

%

\newpage
\appendix
\section{Results obtained with the analysis B}
\label{app:analysisB}

In Sections \ref{sec:neutron} and \ref{sec:electroweak}, the radius of the nuclear neutron distribution and the electroweak mixing angle have been studied using the so-called analysis A of CENNS-10 data~\cite{Akimov:2020pdx}, whose selection criteria allow to put more stringent constraints on the parameters of interest. For completeness, here we present the results obtained using the same fitting procedure developed in Sections~\ref{sec:neutron} and \ref{sec:electroweak} using the data of the CENNS-10 analysis B, in order to check the compatibility and stability of the results. These two different data sets are obtained from the same data-taking campaign and share most of the selection procedure, leading to have most of the data in common. Thus, any attempt to combine the results of analyses A and B in order to obtain a more precise measurement of the physics parameters should be discarded, given the large overlap between the two.

\begin{figure*}[!bth]
\centering
\setlength{\tabcolsep}{0pt}
\begin{tabular}{cc}
\subfigure[]{\label{fig:Rn_AnalysisB}
\begin{tabular}{c}
\includegraphics*[width=0.45\linewidth]{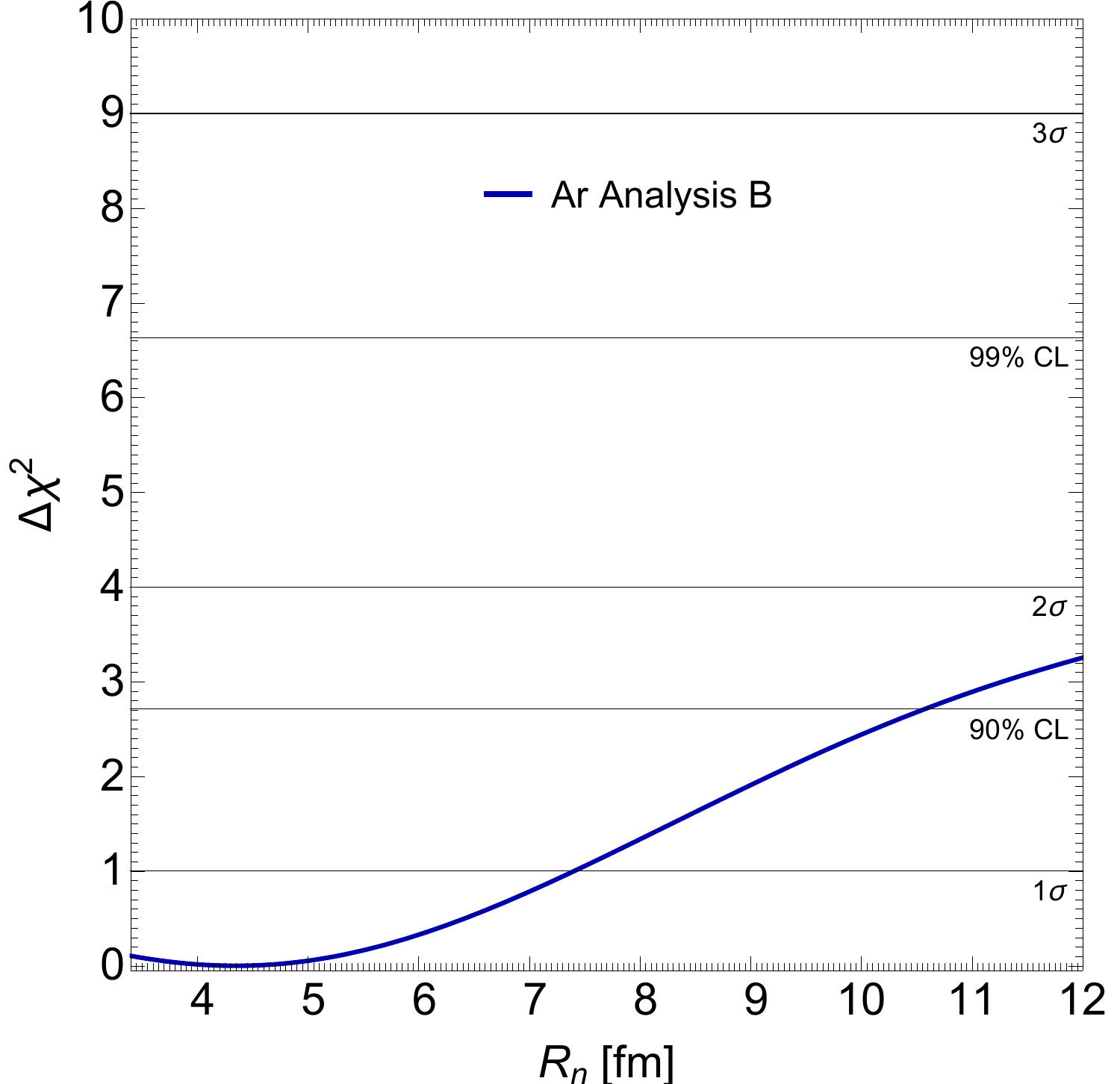}
\\
\end{tabular}
}
&
\subfigure[]{\label{fig:sin2thetaw_AnalysisB}
\begin{tabular}{c}
\includegraphics*[width=0.45\linewidth]{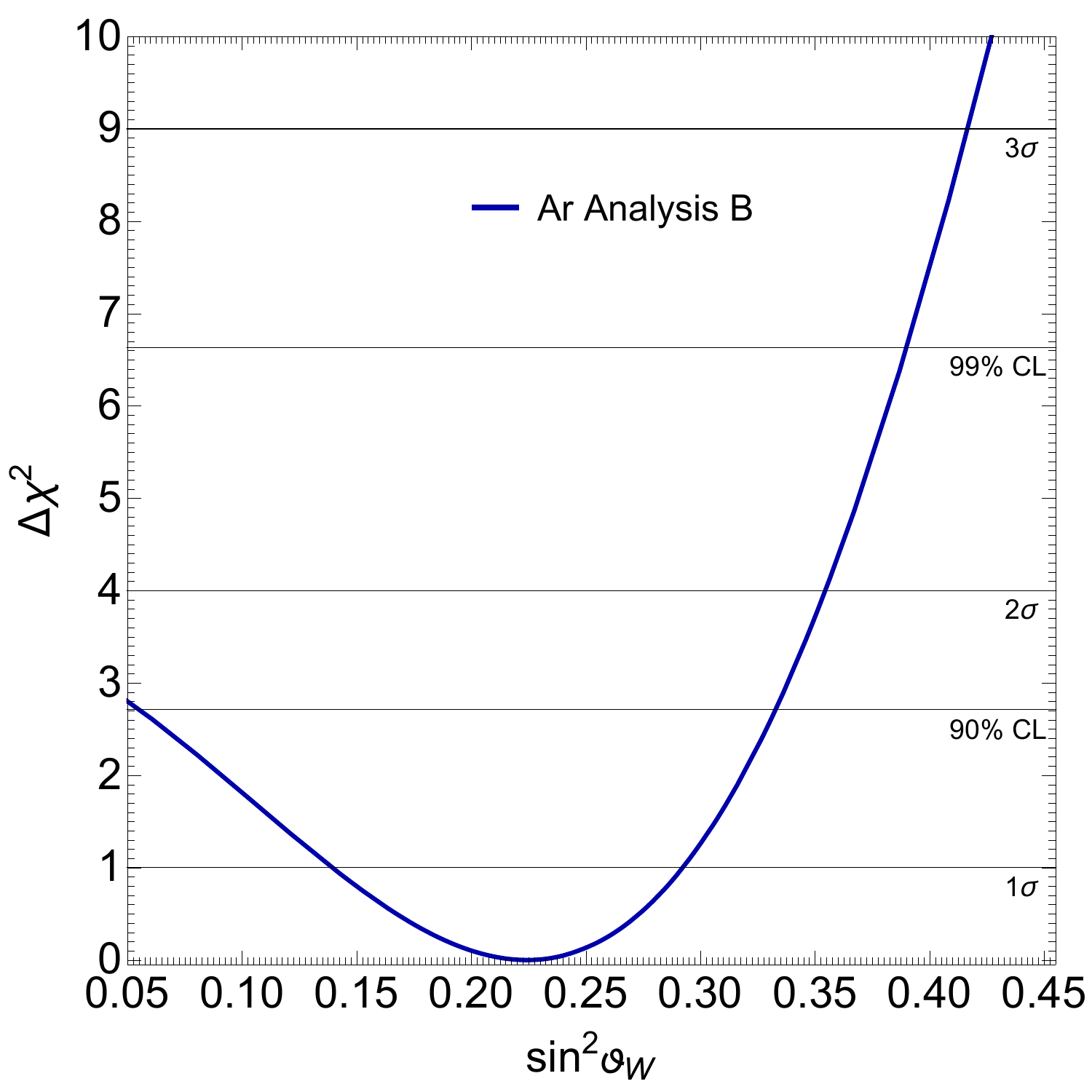}
\\
\end{tabular}
}
\end{tabular}
\caption{$\Delta \chi^2$ profiles for the CENNS-10 analysis B as (a) a function of the neutron rms radius $R_n$, fixing the value of the weak mixing angle $\sin^2\vartheta_W = 0.23587$, and (b) of $\sin^2\vartheta_W$, fixing the value of $R_n$ .}
\end{figure*}

In contrast to analysis A, the slightly different selection in analysis B results into a modified efficiency that is below the efficiency of analysis A except for a small region between 4~$\mathrm{keV}_{ee}$ and 5~$\mathrm{keV}_{ee}$.
In addition, the region of interest of the analysis B is restricted to [4, 30] $\mathrm{keV}_{ee}$, that corresponds roughly to the \cenns signal energy region. The results are presented using 13 bins of 2 $\mathrm{keV}_{ee}$ each. Another difference is that, in analysis B, the delayed component of BRN is not included, thus the background has a single component, $ B^{\mathrm{BRN}}$.\\
The least-squares function becomes
\begin{eqnarray}
\chi^2_{\text{S}}
&=&
\sum_{i=1}^{13}
\left(
\dfrac{
N_{i}^{\text{exp}}
-
\eta_{\mathrm{CE\nu NS}} N_i^{\mathrm{CE\nu NS}}
-
\eta_{\mathrm{BRN}} B_i^{\mathrm{BRN}}}
{\sigma_i}
\right)^2
+
\left( \dfrac{\eta_{\mathrm{CE\nu NS}}-1}{\sigma_{\mathrm{CE\nu NS}}} \right)^2
+
\left( \dfrac{\eta_{\mathrm{BRN}}-1}{\sigma_{\mathrm{BRN}}} \right)^2
,
\label{chi-spectrum-B}
\end{eqnarray}
with
\begin{eqnarray}
\sigma_i^2 = \left( \sigma_i^{\mathrm{exp}} \right)^2 &+&
\left[ \sigma_{\mathrm{BRNES}} B_i^{\mathrm{BRN}}\right]^2,\\
\sigma_{\mathrm{BRNES}} &=& \sqrt{\frac{0.052^2}{13}}=1.4\%,\\
\sigma_{\mathrm{CE\nu NS}} &=& 12\%,\\
\sigma_{\mathrm{BRN}} &=& 14.6\%,\\ \nonumber
\end{eqnarray}
where these quantities have been introduced in Section \ref{sec:neutron}. 
We notice that the systematic uncertainties of analysis B are smaller with respect to analysis A and the energy resolution is better. On the other hand, since the selection performed in analysis B is tighter, the number of expected \cenns events is smaller and the resulting  uncertainty is larger than analysis A.
Fixing the value of the weak mixing angle to the SM one, $\sin^2\vartheta_W = 0.23857$~\cite{Tanabashi:2018oca}, and fitting the radius of the nuclear neutron distribution $R_n$, constraining it to be $R_n > R_p$, we find
\begin{equation}
R_{n}({}^{40}\text{Ar})
<
7.4 
\, (1\sigma)
,
\,
10.55
\, (90\%\, \mathrm{CL})
\, \text{fm}
.
\end{equation}
Figure~\ref{fig:Rn_AnalysisB} shows the
$\Delta\chi^2$
as a function of the neutron rms radius $R_{n}$ for analysis B. Even though a minimum is found at $R_n = 4.36 ~\mathrm{fm}$, the large uncertainty allows only to set limits on the neutron rms radius, which are much weaker than those obtained with the analysis A.

Similarly, we determined the value of $\sin^2\vartheta_W$ fixing $R_n$. This choice does not impact significantly the result, since it has been verified that the extracted value of the weak mixing angle is largely uncorrelated with $R_n$. 
Under these assumptions, we obtain
\begin{equation}
    \sin^2\vartheta_W=0.22^{+0.07}_{-0.09}\,(1\sigma)^{+0.11}_{-0.17}\,(90\,\mathrm{CL}).
\end{equation}
This value is consistent with the SM prediction and with the best fit obtained using analysis A, but a much larger uncertainty is obtained.\footnote{To better judge the consistency one should know the overlap, in terms of number of events, between the two analyses, but this information is missing in Ref.~\cite{Akimov:2020pdx}.} Figure~\ref{fig:sin2thetaw_AnalysisB} shows the
$\Delta\chi^2$
as a function of the weak mixing angle for analysis B.

\end{document}